\documentclass[conference,10pt, compsocconf]{IEEEtran}
\usepackage{mathptmx} % This is Times font
\usepackage{floatrow}
\usepackage{amsmath,amssymb,amsfonts}
\usepackage{algorithmic}
\usepackage{graphicx}
\usepackage{fancyhdr}
\usepackage{textcomp}
\usepackage{multirow}
\usepackage{amsmath}
\usepackage[normalem]{ulem}
\usepackage[hyphens]{url}
\usepackage[sort,compress]{cite}
\usepackage[final]{microtype}
\usepackage[keeplastbox]{flushend}
\usepackage{comment}
\usepackage[table,xcdraw]{xcolor}
\usepackage[font=small, labelfont=bf, textfont=bf]{caption} 
\usepackage{svg}

\usepackage[bookmarks=true,breaklinks=true,letterpaper=true,colorlinks,linkcolor=black,citecolor=blue,urlcolor=black]{hyperref}

\newcommand{\NAME}{NVMExplorer}
\newcommand{\WEBSITE}{\url{http://nvmexplorer.seas.harvard.edu/}}

\title{NVMExplorer: A Framework for Cross-Stack Comparisons of Embedded Non-Volatile Memories
}

\author{\IEEEauthorblockN{Lillian Pentecost\IEEEauthorrefmark{1}\textsuperscript{\textsection}, Alexander Hankin\IEEEauthorrefmark{2}\textsuperscript{\textsection}, Marco Donato\IEEEauthorrefmark{2}, Mark Hempstead\IEEEauthorrefmark{2}, Gu-Yeon Wei\IEEEauthorrefmark{1}, David Brooks\IEEEauthorrefmark{1}}
\IEEEauthorblockA{\IEEEauthorrefmark{1}Harvard University, Cambridge, MA, USA\\
\IEEEauthorrefmark{2}Tufts University, Medford, MA, USA \\
\url{lillian\_pentecost@g.harvard.edu}, \url{alexander.hankin@tufts.edu}}
}

\begin{document}
\maketitle
\begingroup\renewcommand\thefootnote{\textsection}
\footnotetext{Authors contributed equally to this work.}
\endgroup
\thispagestyle{plain}
\pagestyle{plain}

%%%%%% -- PAPER CONTENT STARTS-- %%%%%%%%
\begin{abstract}

The current computing landscape is dominated by data-intensive applications, making data movement one of the most prominent performance bottlenecks. 
With repeated off-chip memory access to DRAM driving up power, and SRAM technology scaling and leakage power limiting the efficiency of embedded memories, there is a need for new memory systems that can enable denser, more energy-efficient future on-chip storage.  
The actively expanding field of emerging, embeddable non-volatile memory (eNVM) technologies is providing many potential candidates to satisfy this need. 
However, eNVM cell technologies are in vastly different stages of development and introduce distinct trade-offs in terms of density, read, write, and reliability characteristics.  

We present NVMExplorer (\WEBSITE): a cross-stack design space exploration framework to compare and evaluate future on-chip memory solutions with system constraints and application-level impacts in-the-loop. 
This work uses NVMExplorer to evaluate eNVM-based storage for a range of application and system contexts including machine learning on the edge, graph analytics, and general purpose cache.  
Additionally, NVMExplorer provides an interactive and easily navigable set of data visualizations, which allow users to quickly answer their specific questions regarding eNVMs, filter according to system and application constraints, and efficiently iterate and refine the design space.
%\footnote{\textcolor{red}{HPCA’s Call for Papers includes ``Measurement, Modeling, Simulation'', and HPCA has a rich history as a venue for tools and frameworks of great importance to the architecture community, particularly when those frameworks empower a broad set of future architectural studies (e.g., NeuroMeter from HPCA 2021 and ``A performance analysis framework for optimizing OpenCL applications on FPGAs'' from HPCA 2016).}}.
\vspace{-5pt}
\end{abstract}

\section{Introduction}

The wide adoption of data-intensive algorithms to tackle today's computational problems introduces new challenges in designing efficient computing systems to support these applications. 
Hardware specialization has shown potential in supporting state-of-the-art machine learning and graph analytics algorithms across several computing platforms; however, data movement remains a major performance and energy bottleneck. 
As repeated memory accesses to off-chip DRAM impose an overwhelming energy cost, we need to rethink the way embedded memory systems are built in order to increase on-chip storage density and energy efficiency beyond what is currently possible with SRAM.

In recent years, CMOS-compatible, embedded nonvolatile memory (eNVM) research has transitioned from articles and technical reports to manufacturing flows and product lines. 
These technologies hold incredible promise toward overcoming the memory wall problem. 
For example, one approach inspired by these new technologies combines the advantages of highly specialized architectures with the benefits of non-volatile memories by leveraging analog compute capabilities~\cite{prime, isaac, pipelayer, cascade}. 
On the other hand, the need for optimized on-chip storage solutions and memory innovation applies both to specialized hardware accelerators and for general-purpose CPU systems as well.
More broadly, prior works have unveiled incredible potential improvements in storage density and energy efficiency by employing eNVMs across various architecture domains~\cite{hankin, deepnvm++, maxnvm}.
With many publications showcasing the benefits of eNVM storage technologies, it is critical for system designers to be able to explore their varying capabilities and empower efficient future on-chip storage.
Unfortunately the architecture and broader research community lacks a holistic tool to quantify the system and application-level implications of memory cell technologies and to make informed decisions while navigating the vast eNVM design space.

\begin{figure}[t]
    \centering
    \floatbox[{\capbeside\thisfloatsetup{capbesideposition={right,bottom},capbesidewidth=0.4\hsize}}]{figure}[\FBwidth]
{\caption{Number of NVM publications from VLSI, ISSCC, and IEDM 2016-2020 (cited in text) shows strong interest in RRAM and STT and emerging technologies, such as ferroelectric-based ones.
%shows sustained interest in RRAM and STT and emerging interest in other technologies, such as ferroelectric-based ones.
}}
{\includegraphics[width=0.99\hsize]{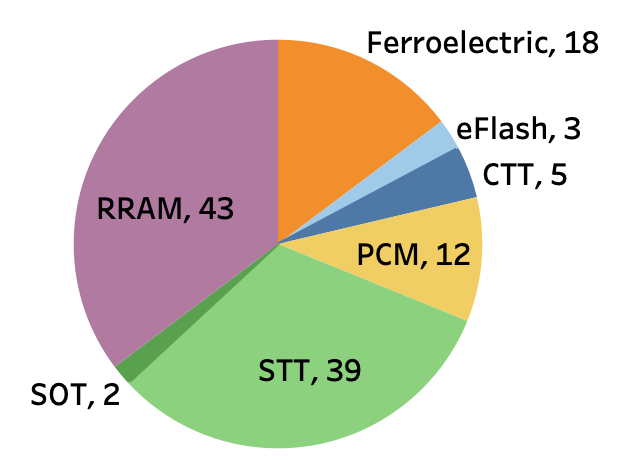}\label{fig:design-space}}
%\vspace{-20pt}
\end{figure}

\begin{figure*}[t]
    \centering
    \includegraphics[width=0.95\hsize]{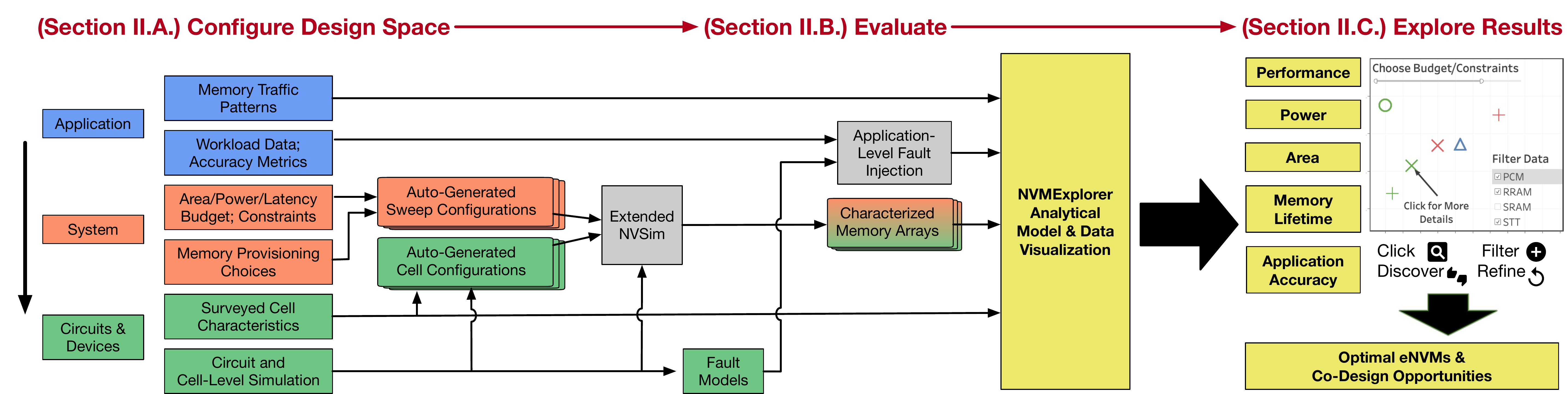}
    \caption{\NAME~framework overview; cross-stack design space specifications and application characteristics are evaluated in an efficient multi-stage process, then displayed in an interactive set of data visualizations to enable informed, application-aware comparisons of future on-chip storage solutions, as described in more detail in Section \ref{sec:nvmexp}.}
    \vspace{-5pt}
    \label{fig:diagram_of_framework}
\end{figure*}

Figure~\ref{fig:design-space} summarizes device and circuit conference publications relating to eNVMs from 2016 to 2020 \cite{2016_1,2016_2,2016_3,2016_4,2016_5,2016_6,2016_7,2016_8,2016_9,2016_10,2016_11,2016_12,2016_13,2016_14,2016_15,2016_16,2016_17,2016_18,2016_19,2016_20,2016_21,2016_22,2016_23,2016_24,2016_25,2016_26,2016_27,2016_28,2016_29,2016_30,2016_31,2016_32,2016_33,2016_34,2016_35,2016_36,2016_37,2016_38,2017_1,2017_2,2017_3,2017_4,2017_5,2017_6,2017_7,2017_8,2017_9,2017_10,2017_11,2017_12,2017_13,2017_14,2017_15,2017_16,2017_17,2017_18,2017_19,2017_20,2017_21,2017_22,2017_23,2017_24,2017_25,2018_1,2018_2,2018_3,2018_4,2018_5,2018_6,2018_7,2018_8,2018_9,2018_10,2018_11,2018_12,2018_13,2018_14,2018_15,2018_16,2018_17,2018_18,2018_19,2018_20,2019_1,2019_2,2019_4,2019_5,2019_6,2019_7,2019_8,2019_9,2019_10,2019_11,2019_12,2019_13,2019_14,2019_15,2019_16,2019_17,2019_20,2019_21,2019_22, 2020_1,2020_2,2020_3,2020_4,2020_5,2020_6,2020_7,2020_8,2020_9,2020_10,2020_11,2020_12,2020_13,2020_14,2020_15,2020_16,2020_17,2020_18,2020_19,2020_20,2020_21,2020_22,2020_23,2020_24,2020_25}.
In the past five years, consistent interest in RRAM and STT was accompanied by emerging solutions with different physical properties such as FeFET-based memories. 
Each published example offers compelling and distinct trade-offs in terms of read and write characteristics, storage density, and reliability.  
In addition, the space of eNVM technologies is constantly evolving with certain technologies moving out of fashion or into production. 
Given the fluidity and complexity of this design space, application experts and system designers need to be able to evaluate which cell technologies are most likely to provide better efficiency, higher storage density, or improvements on other key metrics in the context of different computing demands. 
Similarly, device designers and memory architects need high-level guidance to co-design their innovations toward more practical and maximally beneficial future, heterogeneous memory systems.

This work introduces \NAME, an end-to-end design space exploration framework that addresses key cross-stack design questions and reveals future opportunities across eNVM technologies under realistic system-level constraints, while providing a flexible interface to empower further investigations. 
In this work, we describe \NAME\ and present case studies made uniquely possible by the capabilities of \NAME. In summary, \NAME\ makes the following key contributions:
\begin{itemize}
    \item An open-source code base including: 
    \begin{itemize}
        \item A database of eNVM cells described in recent literature (122 surveyed ISSCC, IEDM, and VLSI publications) (Section \ref{subsec:cells})
        \item  A ``tentpole'' methodology to summarize limits and trends across technology classes (Section~\ref{subsec:tentpoles})
        \item Our end-to-end evaluation flow (Fig. \ref{fig:diagram_of_framework})
        \item Extensive source-code documentation
        \item Many example configuration files and tutorial materials for cross-stack design studies
        \item An interactive web-based data visualization dashboard (Section \ref{subsec:visualization})
    \end{itemize} 
    \item A unified platform to explore the viability of eNVMs in specific application and system settings, which reveals cross-stack dependencies and optimization opportunities, in addition to reproducing and expanding previous published studies, (e.g., \cite{maxnvm} \cite{hankin}) (Section \ref{sec:case_studies}).
    \item A unified platform to perform co-design studies of application properties, system constraints, and devices in order to bridge the gap between architects and device designers for future eNVM solutions. Our example co-design studies reveal both opportunities and potential disconnects among current research efforts (Section \ref{sec:codesign}).
\end{itemize}

After describing \NAME\ (Section \ref{sec:nvmexp}), we present a snapshot of the current eNVM landscape and extract a representative range of cell-level behavior (Section \ref{sec:tech_landscape}).
Surveying recent eNVM publications reveals diverse characteristics, highlighting the challenge in identifying solutions that satisfy a broad range of application scenarios.
Thus, Section \ref{sec:case_studies} presents application-driven case studies using \NAME\ to explore and analyze eNVM storage solutions for DNN inference acceleration, graph processing, and general-purpose compute.
We find that each eNVM is viable in certain contexts, and the most compelling eNVM is dependent on application behavior, system constraints, and device-level choices. 
This finding suggests the existence of many possible architecture-device co-design opportunities, which is the focus of Section \ref{sec:codesign}. 
%We consider a set of both architecture-driven and device-driven co-design opportunities including: using an emergent, alternative FeFET cell for graph processing; relaxing cell-level parameters to optimize for application or system constraints; calibrating advantages of multi-level programming per technology; and write-buffering. 
%Interestingly, our co-design studies reveal both opportunities and potential disconnects between the priorities of architects and device-designers in maximizing memory efficiency in a realistic system context. 
Finally, we differentiate \NAME\ from related tools (Section \ref{sec:relatedwork}).

\begin{figure*}[t]
    \centering
    \includegraphics[width=0.95\textwidth]{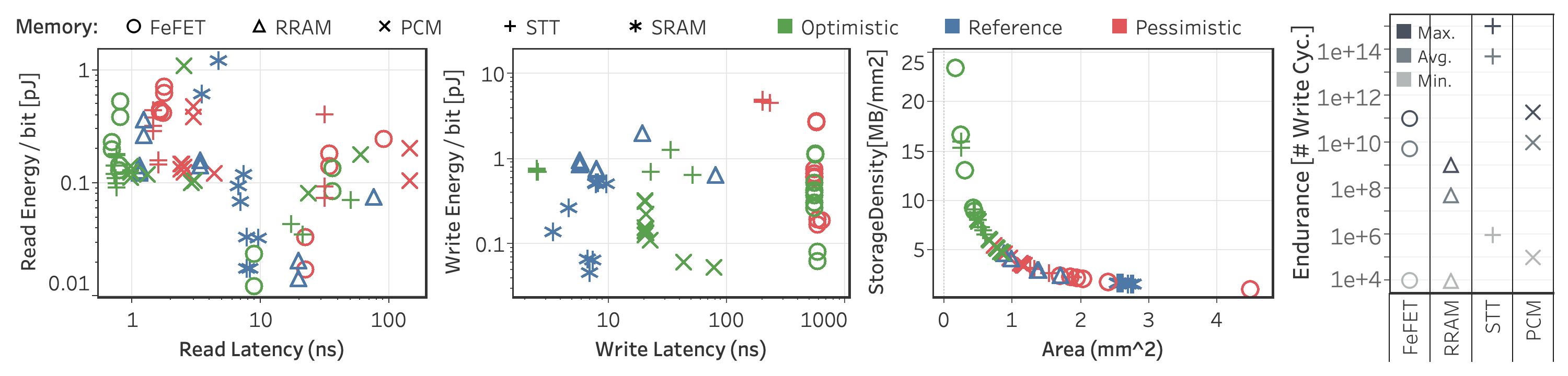}
    \caption{For fixed capacity (4MB) and under various optimization targets, array-level metrics reveal each eNVM has unique, compelling attributes. Note Pessimistic PCM write latency ($>10\mu$s) is omitted for clarity in the write energy vs. latency plot.}
    \vspace{-10pt}
    \label{fig:motivate_area_readperf}
\end{figure*}

\section{\NAME}
\label{sec:nvmexp}
%In this section, we present an overview of our tool. 
At a high level, \NAME~is a comprehensive design space exploration (DSE) framework integrating application-level characteristics, system constraints, and circuit and device parameters in a publicly-available, simple-to-use flow. 
The overall structure of \NAME\ (Fig.~\ref{fig:diagram_of_framework}) relies on three stages, described in more details in the following subsections: 
%\textcolor{red}{Text that follows here need to be refreshed to follow the overview fig more closely?}
%\noindent \NAME~has three major components:
\begin{enumerate}
    \item A comprehensive cross-stack configuration interface to specify the design space of interest. This configuration spans the computing stack from application (blue) and system (orange) down to circuits and devices (green). 
%    \item First, it relies on a wide survey of eNVM cell designs and low-level properties to establish and explore the breadth and key characteristics of memory proposals
%    \item Next, it provides an extensible and modular user interface to specify end-to-end optimization goals, system constraints such as area or power budget, application details such as traffic patterns or bandwidth requirements, or alternative technology cell characteristics
    \item An evaluation engine which automatically generates configurations, simulates memory arrays, processes application behavior, computes key metrics such as performance, power, area, accuracy, and lifetime, and generates meaningful visualizations. Evaluation steps which extend existing tools are shaded grey in Fig. \ref{fig:diagram_of_framework}.
    %\begin{enumerate}
    %    \item First, a modified and extended version of NVSim extrapolates cell characteristics to memory array-level metrics
    %    \item Then, our analytical model estimates overall dynamic power, latency, and memory lifetime as dictated by application use-case and system settings
    %    \item Additionally, application data and memory fault models can be connected to an application-level fault injection tool to quantify the impacts of memory faults
    %\end{enumerate}
    \item An interactive, web-based visualization tool to aide discovering, filtering and refining eNVM design points.
\end{enumerate}

\vspace{-5pt}
\subsection{Cross-Stack Configuration} 
\label{subsec:cross-stack}
%and Constraints}
To evaluate and compare eNVM solutions in system settings, it is not just cell or even array-level characteristics of a particular technology that matter.
Rather, viable solutions depend on the area/power budget of a system and how applications running on that system interact with the memory. 
\NAME\ provides a rich interface for configuring key application, system, and circuit and device parameters.

At the application level, the user inputs information about memory traffic, which may include the number of read and write operations, their proportion relative to the total number of memory accesses, and how accesses are spread out over execution time. 
These configuration parameters may be fixed values (e.g., characterization results of a specific workload) or provided as ranges to generate generic memory traffic patterns.
%With either type of application input, NVMExplorer will analytically evaluate a given memory specification over a range of application scenarios. 
Some applications may have additional demands or metrics which are tightly related to memory technology characteristics. 
For example, machine learning applications or approximate computing methods may trade-off relaxed accuracy for performance and energy, and \NAME~also provides an interface for designers to study the application interactions and implications of fault-prone eNVM solutions.  

At the system level, the user has the freedom to evaluate a wide variety of memory configuration options by either setting performance, power, and area constraints and optimization goals or by choosing memory array specifications such as capacity, multi-level programming, bank configuration, and more. 
The circuits and devices level of the design space configuration comprises per-technology memory cell characteristics, in addition to sensing and programming circuitry choices. 
\NAME~also provides a database of eNVM cell configurations derived from ISSCC, IEDM, and VLSI publications, as described in Section \ref{sec:tech_landscape}, but it is also possible (and encouraged!) for users to extend the current \NAME~database with new simulation-based (\textit{i.e.} SPICE or TCAD models), measured, or projected circuit and device properties.
Once the full-stack specifications are set, \NAME~automatically generates configuration files, which are used as input to the evaluation engine.

\begin{table*}[t] \small \center
\caption{High-level listing of memory cell technologies and ranges for key characteristics; recent publications are complemented by simulation and industry references to form technology cell definitions discussed in Section \ref{subsec:cells}.} 
\resizebox{0.7\textwidth}{!}{
\begin{tabular}[b]{c|c|c|c|c|c|c|c|c|c|}

\cline{2-9}
                                                   & \textbf{SRAM} & \textbf{PCM}             & \textbf{STT}        & \textbf{SOT}                                      & \textbf{RRAM}      & \textbf{CTT} & \textbf{FeRAM}           & \textbf{FeFET}           \\ \hline
\multicolumn{1}{|l|}{Cell Area [$F^2$]}           & 146           & 25-40                    & 14-75               & \cellcolor[HTML]{C0C0C0}   [20]                      & 4-53               & 1-12         & \cellcolor[HTML]{C0C0C0} & 4-103                    \\ \hline
\multicolumn{1}{|l|}{Tech. Node [$nm$]}         & 7-16          & 28-120                  & 22-90               & \cellcolor[HTML]{C0C0C0} [1000]                              & 16-130             & 14-16        & 40                       & 45                       \\ \hline
\multicolumn{1}{|l|}{MLC}                         & no            & yes                      & yes                 & yes                                               & yes      & yes          & yes                      & yes                     \\ \hline
\multicolumn{1}{|l|}{Read Latency {[$ns$]}}          & 0.5-1.5       & \cellcolor[HTML]{C0C0C0}  [1-100]         & 1.3-19              & 1.4-11                                              & 3.3-2e3            &   \cellcolor[HTML]{C0C0C0}           & 14                       & \cellcolor[HTML]{C0C0C0} \\ \hline
\multicolumn{1}{|l|}{Write Latency {[$ns$]}}         & 0.5-1.5       & 10-3e4                   & 2-200               & 0.35-17                                        & 5-1e5              & 6e7-2.6e9    & 14-1e3                   & 0.93-1.3e3             \\ \hline
\multicolumn{1}{|l|}{Read Energy {[$pJ$]}}         & 1.1-2.4      & \cellcolor[HTML]{C0C0C0}        & 0.21-1.2            & \cellcolor[HTML]{C0C0C0}                          & 1e-3               &   \cellcolor[HTML]{C0C0C0}            & 0.001                    & \cellcolor[HTML]{C0C0C0} \\ \hline
\multicolumn{1}{|l|}{Write Energy {[$pJ$]}}        & 1.1-33        & \cellcolor[HTML]{C0C0C0}              & 0.6-4.5             &  \cellcolor[HTML]{C0C0C0} [0.015-8]                                                & 0.68               &     \cellcolor[HTML]{C0C0C0}          & \cellcolor[HTML]{C0C0C0} & 0.0003-0.01     \\ \hline
\multicolumn{1}{|l|}{Endurance [Cycles]} & N/A           & 10$^{5}$--10$^{11}$      & 10$^{5}$--10$^{15}$ & \cellcolor[HTML]{C0C0C0}                          & 10$^{3}$--10$^{8}$ & 10$^{4}$     & 10$^{4}$--10$^{11}$      & 10$^{7}$-10$^{11}$                  \\ \hline
\multicolumn{1}{|l|}{Retention [s]}            & N/A           & 10$^{8}$--10$^{10}$      & 10$^{8}$            & 10$^{8}$                                          & 10$^{3}$--10$^{8}$ & 10$^{8}$     & 10$^{5}$--10$^{8}$       & \cellcolor[HTML]{C0C0C0}  \\ \hline
%\multicolumn{1}{|l|}{error rate}                   & N/A           & \cellcolor[HTML]{C0C0C0} & 10$^{4}$--10$^{11}$ & \cellcolor[HTML]{C0C0C0}{\color[HTML]{C0C0C0} ??} & 10$^{10}$          &              & \cellcolor[HTML]{C0C0C0} & \cellcolor[HTML]{C0C0C0} &                 \\ \hline
\end{tabular}
}
\centering
\includegraphics[width=0.19\textwidth]{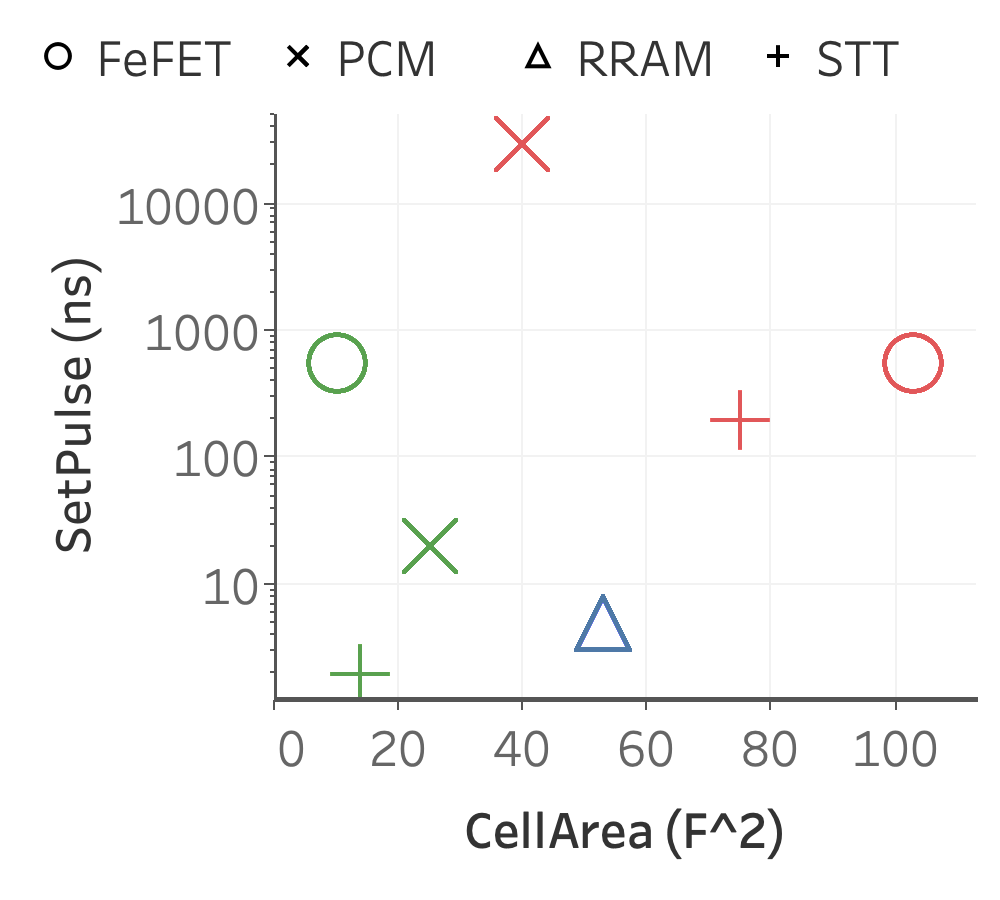}

\vspace{-5pt}

\label{tab:cell_level_ranges}
\end{table*}

\vspace{-5pt}

\subsection{Evaluation Engine}

%\NAME~natively supports a wide range of design space configurations including a wide array of application choices and generic memory traffic patterns, system constraints, and eNVM cell characteristics. 
%However, the user is also free to configure their own design space. 
Given the auto-generated cell and system-level sweep configurations, the evaluation engine produces memory array architecture characterizations and computes application- and system-level power, performance, area, and reliability metrics. 
\NAME\ combines a customized memory array simulator, an application-level fault injection tool, and an analytical model to extrapolate application-level metrics. 

To characterize memory arrays, we rely on a customized version of NVSim, a previously validated tool to compute array-level timing, energy, and area~\cite{NVSim}. 
We build on existing efforts to extend NVSim to support multi-level cells and ferroelectric-based eNVMs \cite{fefet-arxiv, maxnvm}. 
In addition, we modified the tool interface to ease data collection and post-processing. 
We introduce the capabilities of \NAME~in comparing eNVMs in Section~\ref{subsubsec:mem_array_comp}. 
Results of cell-level and circuit-level simulations can be used to parameterize fault models and perform application-level fault injection, as described in Section~\ref{subsubsec:fault_modeling}.
For performance estimations, in lieu of cycle-accurate simulation, we utilize a long-pole, bandwidth driven model that takes memory access latency and available read/write bandwidth and compares aggregated access latency per workload execution and per second of execution to workload access statistics.
This is similar in spirit to performance models in \cite{dark_silicon, nvdla}, and it serves the primary purpose of identifying memory solutions that cause application slowdown, rather than predicting precise latency implications.
To extract other critical application-level metrics, such as energy, we aggregate the read and write access energy based on the number of application accesses and array energy-per-access with the leakage power, scaling according to use-case and wake-up frequency for intermittent operation.
Similarly, memory lifetime is extrapolated by comparing the average reported endurance to the write access pattern per workload and the use-case.

\subsubsection{Example Array-Level Comparison}
\label{subsubsec:mem_array_comp}

Figure~\ref{fig:motivate_area_readperf} presents example array characterization output generated by \NAME~after evaluating various eNVM configurations implemented in a 22nm node. 
The design points are color-coded to highlight optimistic (green), pessimistic (red), or reference (blue) designs across surveyed publications per cell technology. 
The figure also reports the characteristics of 16nm SRAM as a comparison point. 
For each technology, we show array characterization under different optimization goals, which result in a variety of internal array architectures. % (e.g., bank arrangement and column muxing for read-out).
For example, we observe a wide range for the read-energy-per-bit of an iso-capacity SRAM array. 
This result reflects the effect of different array optimization targets (read energy-delay product, write characteristics, area) on the internal bank configuration and periphery overhead.

%As an example of the capabilities of \NAME, Figure~\ref{fig:motivate_area_readperf} presents a look at various iso-capacity array characteristics for a set of eNVM technology classes that are promising for adoption today. 
%The coloring of these options refers to optimistic (green), pessimistic (red), or reference (blue) underlying cell properties per technology (described more in Section~\ref{subsec:tentpoles}) at a 22nm technology node, and we include 16nm SRAM as a comparison point. 
%We note that for each of these technologies, we show array characterization under different optimization targets, which result in a variety of internal array architectures (e.g., bank arrangement and column muxing for read-out).
%For example, the range in read-energy-per-bit for iso-capacity SRAM points in Figure \ref{fig:motivate_area_readperf} reflects arrays optimized for read latency at 1 pJ/bit access due to many internal banks and resulting periphery overhead for efficient access compared to arrays optimized for latency, area, or write characteristics that result in different array layouts.

This preliminary study already provides a few key takeaways. 
Each eNVM is able to attain read access characteristics competitive with SRAM, with the exception of an array characterized with pessimistic underlying PCM cell characteristics. 
However, write access characteristics vary dramatically across published eNVM examples, in addition to the range of reported endurance per technology.
The tension between these properties and potential storage density (even in the absence of multi-level cell programming) indicates that array-level comparison in isolation may guide a system designer towards sub-optimal solutions. 
For example, a FeFET-based memory may seem a fitting choice for high-density, read-performant storage, but we find that both performance and energy efficiency of those memories are highly shaped by application traffic patterns and underlying cell assumptions. 
Thus, the cross-stack nature of data exploration supported by \NAME~is essential in guiding system-level choices and further investigation.

\subsubsection{Fault Modeling and Reliability Studies} \label{subsubsec:fault_modeling}

In addition to characterizing memory performance, power, area, and lifetime, \NAME\ extends previously validated efforts in application-level fault injection to provide an interface for fault modeling and reliability studies~\cite{ares}.
Users can provide an expected error rate or more detailed, technology-specific fault models and storage formats to perform fault injection trials on application data stored in different eNVMs. 
To quantify the impact on application-specific metrics of accuracy, the fault injection tool is tightly integrated with application libraries for data-intensive workloads, including PyTorch for DNNs and snap for graph processing \cite{pytorch, snapnets}, as well as numpy for generic application data.
As a demonstration, we perform SPICE simulation and extract fault charactieristics associated with single-level vs. multi-level cell (SLC vs. MLC) programming and sensing circuitry characteristics. 
In this work, we consider a subset of eNVMs, namely, RRAM, CTT, and FeFET, whose fault characteristics could be derived from existing modeling efforts \cite{maxnvm, fefet-arxiv}.
We use our extended fault injection framework to simulate the impact of storing workload data in SLCs vs. MLCs in Section \ref{subsec:mlc}.
Armed with these additional capabilities, \NAME\ can replicate the results of previous considerations of eNVM storage reliability \cite{maxnvm}, in addition to providing a broader platform for studying the interactions between programming choices, cell characteristics, and application accuracy.

%\textcolor{red}{Describe application-level fault injection and how this complements other high-level metrics}

%This is good, but I think we can make this point in previous para
%This analysis raises a lot of questions about the trade-offs hiding within the choice of technology proposal. 
%The key characteristics and limiting factors of NVM technologies, plus the application-level impact, will vary, and therefore we need \NAME to expose and explore these memory options and configurations. 
%In the subsequent sections, we demonstrate more of the capabilities of our tool and how the information gleaned can be used by every corner of the architecture community. 

\vspace{-5pt}
\subsection{Exploring Results \& Conducting Studies} \label{subsec:visualization}

%\subsubsection{Data Visualization Tool}

The figures in this work are snapshots from \NAME's interactive web-based data visualization tool, which will be freely available at the time of publication of this work.
In each study, we filter and constrain evaluated results according to system optimization priorities and application use cases, as described in the text. 
The basic \NAME\ data visualization dashboard presents power, performance, area, and memory lifetime results across all user-configured sweep results (e.g., many application traffic patterns, array provisioning choices, and/or eNVM cell configurations) alongside array-level metrics for a holistic design exploration experience.
A user can filter results in terms of important constraints (e.g., latency or accuracy targets, power or memory area budget) and identify design points of interest. 
While several features of these visualizations, built using Tableau~\cite{tableau}, are evident in the figures in this work, including dynamic filtering across plots, click-and-drag to narrow the design space, and pop-up details about results, we encourage the reader to use their imagination in how they might explore and filter the data shown in alternative ways according to their interests, questions, or confusions.

\section{Technology Landscape} \label{sec:tech_landscape}

\NAME\ provides a broad survey of published eNVM examples (Section \ref{subsec:cells}), which can be parameterized so that systems experts can make meaningful, high-level comparisons across technologies despite different underlying trade-offs and maturity (Section \ref{subsec:tentpoles}). We validate this approach per-technology against fabricated memory arrays (Section \ref{subsec:validation}).

\subsection{Cell Definitions} \label{subsec:cells}

We compile device- and array-level data across eNVM technologies, as summarized in Table \ref{tab:cell_level_ranges} alongside SRAM properties. 
We source the majority of the cell-level parameters from ISSCC, IEDM, and VLSI publications and focus primarily on works from 2017-2020 to reflect the most recent range of achievable behavior per technology. 
Previous efforts detailed the physical properties and limitations per technology \cite{ibm_survey}, while \NAME\ focuses on compiling sufficient cell-level details and leaning on existing technology models to provide a broad and practical database of cell definitions. 
While we hope these extracted cell definitions are helpful to the community in calibrating the current state-of-the-art, \NAME\ is extensible as the design space continues to evolve, as demonstrated in Section \ref{sec:codesign}.

%We also include a couple older technology proposals for reference. 
%The full list of parameters includes: Capacity [Mb],	Access Type (e.g., Xbar),	Material, 	Structure (e.g., 1T1R),	Cell Area [F2],	Cell area, [um2],	Fab Process [nm],	Resistance Ratio,	R High [ohm],	R Low [ohm],	Reset Energy [nJ],	Read Energy [nJ],	Endurance,	Retention [s],	Temp for Retention [C],	Leakage Current [pA],	Leakage Power [W],	Read Voltage [V],	Read Speed [ns],	SET Speed [ns],	RESET current [uA],	RESET Speed [ns],	SET Current [uA],	SET Voltage [V],	RESET Voltage [V],	Error Rate,	Multi-Level. Parameter ranges for a subset of parameters is shown for illustration purposes in Table~\ref{tab:cell_level_ranges}.

%Table \ref{tab:cell_level_ranges}, we can observe some key characteristics that an architect will care about for a memory technology, including the range of demonstrated cell area [F$^{2}$], fab process [nm], endurance [write-cycles], and data retention [s]. 
The technology classes in Table \ref{tab:cell_level_ranges} are at different levels of maturity. 
For example, SOT is a relatively recent technology, and while it boasts very impressive write speed and lower write current compared to STT, it is not yet published at advanced process nodes. 
We also see that endurance varies by multiple orders of magnitude across different technologies. 
Thus, adoption will depend on the write intensity of target applications and system dynamics, so incorporating memory lifetime estimation becomes a critical design consideration.

Grey cells in Table \ref{tab:cell_level_ranges} indicate parameters unavailable in recent publications. 
This could be for reasons of propriety from industry fabrication or experimental constraints. 
However, for architects, it is important to have some concept of the possible range of values associated with these parameters. 
In those cases where SPICE models for a technology are available, we use simulations to fill in missing parameters. 
Alternatively, we consider older publications and consult with device experts to reason about cell and array parameters. 
%This includes filling in missing parameters with either optimistic or pessimistic assumptions about achievable cell characteristics to establish reasonable bounds of the design space per technology, as described in Section \ref{subsec:tentpoles}.

\subsection{Tentpoles of the Design Space} 
\label{subsec:tentpoles}
Comparing eNVMs at varying stages of development and with varying underlying physical properties is a challenging task. 
The case studies in this work aim to provide high-level guidance and relative judgments about which eNVM cell technologies are worthy of further investigation under specific system and application constraints.
Thus, rather than focus in on specific, physically accurate cell configurations, we aim to model the bounds of what is conceivable per eNVM technology across the full range of published recent academic work.
We liken identifying and evaluating these bounds per-technology to forming the poles of a tent that encompasses the full extent of eNVM properties, so we call the extrema in terms of cell-level characteristics (i.e., smallest, lowest read energy, best retention vs. largest cell size, lowest endurance) the device-level ``tentpoles''. 
In an actively evolving technology space, this approach allows us to make meaningful classifications about which technologies are potentially adoptable solutions.
These modeling choices are classified into two fixed cell configurations for applicable technologies, as summarized in Section \ref{subsec:opt_pess} and the figure alongside Table \ref{tab:cell_level_ranges}.
We validate that the ``tentpoles'' of the cell-level design space result in array-level characterization that provides coverage of published memory array properties, as discussed in Section \ref{subsec:validation}. 

\subsubsection{Optimistic and Pessimistic Cell Configurations}
\label{subsec:opt_pess}

For the technology classes most represented in our survey (Fig. \ref{fig:design-space}), we compute which published example has the best-case and worst-case storage density in terms of Mb/F$^{2}$, and this data serves as the foundation of the bounds of the cell-level design space; those points which are most and least dense across recent published examples. 
Any critical cell-level parameters not reported with those cell definitions are assigned values (e.g., read characteristics and programming settings) using the best (lowest power, highest efficiency) or worst (highest power, lowest efficiency) value per metric across all other recent publications with sufficient supporting data. 
These best-case and worst-case technologies per class form the tentpoles of the underlying cell design space, and we label these fixed cell definitions as ``optimistic'' or ``pessimistic'' accordingly.
For the purposes of the case studies presented in Sections \ref{sec:case_studies} and \ref{sec:codesign}, all array- and application-level results are produced using these fixed underlying optimistic and pessimistic cell properties, though we note that a user of \NAME\ can draw either on these constructed, bounding example cells or on the full database of surveyed configurations, or on fully customized definitions with respect to cell size, access properties, and operating conditions (e.g.,read/write voltage, temperature).
Corresponding fault models and error rates for reliability studies are extracted after optimistic vs. pessimistic cell-level properties are fixed, as discussed in one of the presented case studies (Section \ref{subsec:mlc}).

This approach helpful for many reasons: for one, these extremes help us answer exploratory questions about what we will likely see in the near future; secondly, comparing the best-case of one technology to the worst-case of another can help gauge less mature technologies against more mature reference points; thirdly, if such optimistic configurations are untenable or even pessimistic configurations are attractive in a specific system setting, we can build confidence for further exploration and more detailed modeling efforts without implementing and attempting to meaningfully compare many many cell definitions with insufficient data. 
A limitation of this methodology is that inherent trade-offs between certain parameters for a technology may not be linked (e.g., area, latency, and retention for STT); however, this amalgam of cell properties represent the full spectrum of achievable characteristics per technology, rather than specific fabricated results. 
As a point of additional comparison, the results shown in the following studies include a reference cell configuration for RRAM as a relatively mature eNVM, with parameters derived from a specific industry result \cite{2018_1}. 
The resulting optimistic, pessimistic, and reference cell size and write pulse are shown to the right of Table \ref{tab:cell_level_ranges}.

\begin{figure}[t]
    \centering
    \includegraphics[width=0.98\textwidth]{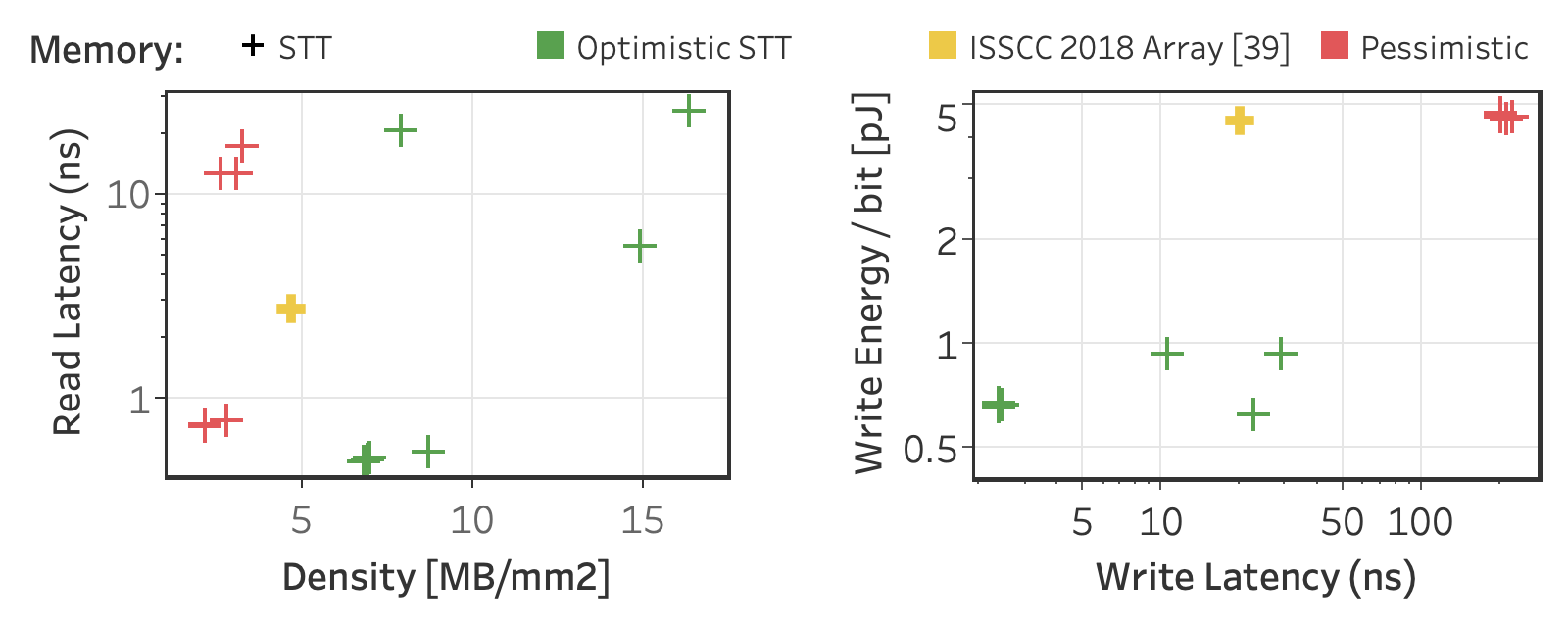}
    \caption{``Tentpole'' STT vs. published array data shows coverage of the space across critical metrics.}
    \label{fig:validate}
    \vspace{-10pt}
\end{figure}

\vspace{-5pt}

\subsection{Validation} \label{subsec:validation}

Our array-level area, energy, and latency characterizations rely on the previously-validated procedures of NVSim to extrapolate cell-level configurations and array design constraints to optimized memory layouts and properties~\cite{NVSim}. 
However, in employing our ``tentpole'' approach, it is critical that we verify that array-level results using our optimistic and pessimistic underlying cell characteristics fully cover and match expectations of existing fabricated eNVM solutions.

Whenever possible, we select publications with array-level characterizations for a given technology, and compare those results to iso-capacity memory arrays modeled through our ``tentpole’’ approach. Figure \ref{fig:validate} shows an example of such an exercise. We compare a 1MB STT-RAM array published at ISSCC in 2018 to optimistic and pessimistic STT design points produced by~\NAME.  Here, we note that our tentpole results effectively represent the range of actual array properties by producing metrics that are both higher and lower, but similar in magnitude, to the reference STT-RAM array.  The studies presented in this work consider only validated configurations for which we were able to either complete this validation exercise or run SPICE-level simulations. 
It is worth noting that \NAME\ is set up to evaluate all cell technologies in Table \ref{tab:cell_level_ranges} (e.g., though SOT is a compelling emerging solution and NVMExplorer users can configure and evaluate SOT-RAM, our survey found insufficient array-level data for validation, so it is omitted in Section \ref{sec:case_studies} and \ref{sec:codesign}). System validation and application characteristics are derived from existing, state-of-the-art references, as addressed in each study in Section \ref{sec:case_studies}. %rigorous validation or future publications 

\section{Application-Driven Case Studies} 
\label{sec:case_studies}
%So far we have looked at how the NVM technology cell survey is done and how this can be used to perform interesting analyses like the aforementioned tentpole study. We now take this one step further and look at 

We now present three case studies that highlight different ways \NAME~can search design spaces in order to identify benefits and limitations of the diverse range of eNVM storage solutions. 
Each scenario presents unique optimization goals and system priorities and, in each case, we compare how each eNVM's power, performance, and area fairs relative to similarly-provisioned SRAM or DRAM in a baseline system.

\begin{figure}[t]
    \centering
    \includegraphics[width=0.95\textwidth]{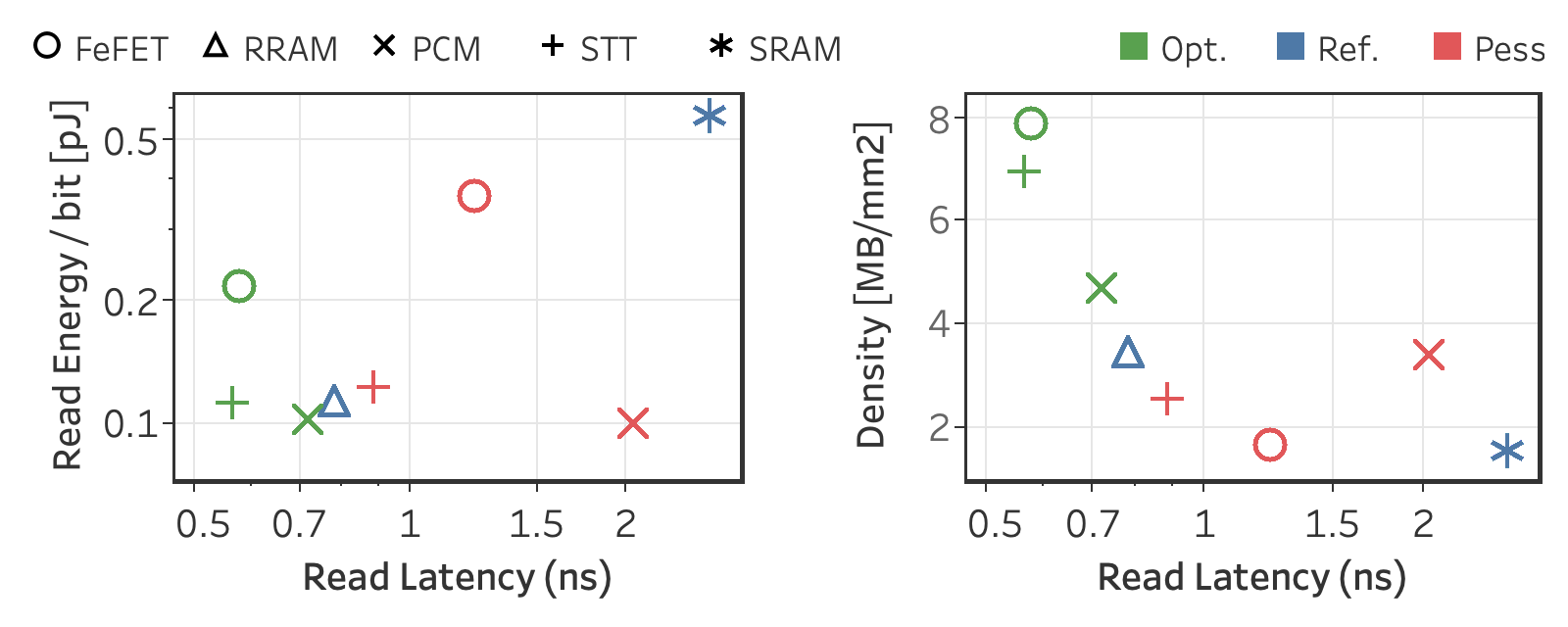}
    \caption{Read characteristics and storage density for 2MB arrays, provisioned to replace on-chip SRAM for NVDLA.}
    \label{fig:dnn_array}
    \vspace{-10pt}
\end{figure}

\begin{figure}[t]
    \centering
    \includegraphics[width=0.48\textwidth]{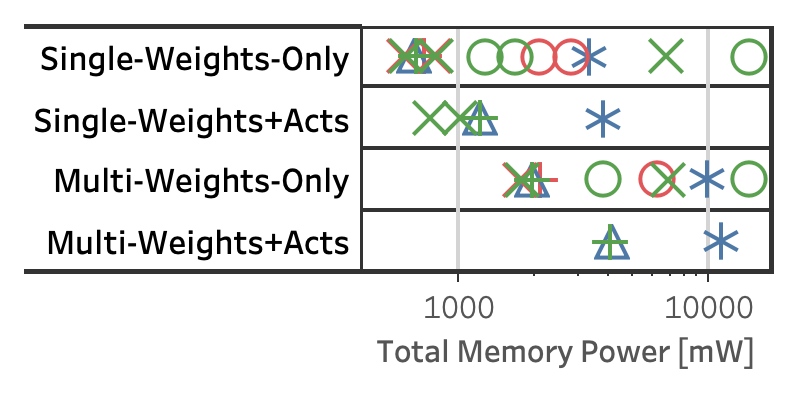}
    \includegraphics[width=0.48\textwidth]{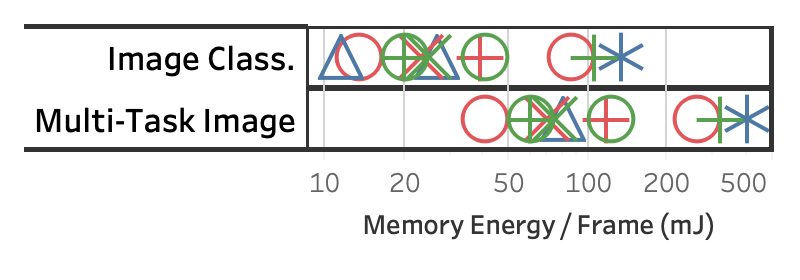}
    \caption{The most energy-efficient eNVM varies under different DNN inference use cases, such as continuous (left, operating power) vs. intermittent (right, reporting energy per input image frame); these results exclude eNVM solutions that are unable to meet application latency and accuracy targets.}
    \vspace{-10pt}
    \label{fig:dnn_app}
\end{figure}

\vspace{-5pt}
\subsection{DNN Inference Accelerator} 
\label{subsec:dnn}
%\textcolor{red}{TODO This will include some limited resilience studies to flex that we have a parallel fault analysis tool}

Prior studies have demonstrated the potential benefits of eNVM storage for Deep Neural Network (DNN) inference accelerators \cite{maxnvm, memti, rram-dnn-blaauw}, albeit with limited scope in terms of eNVM technologies and cross-stack parameters considered. 
\NAME\ empowers researchers to approach a broader set of questions that compare eNVMs in different storage scenarios (e.g., limited to weights vs. storage of DNN parameters and intermediate results) and system constraints (e.g., strict area budget, or power budget).
In this work, we consider two distinct use cases for a DNN inference accelerator:  continuous operation, as in image processing per frame of a streamed video input, and intermittent operation, where the system is woken up per inference task and can leverage the non-volatility of eNVM by retaining DNN parameters on-chip in power-off state between inferences.

\begin{figure}[t]
    \centering
    \includegraphics[width=0.47\textwidth]{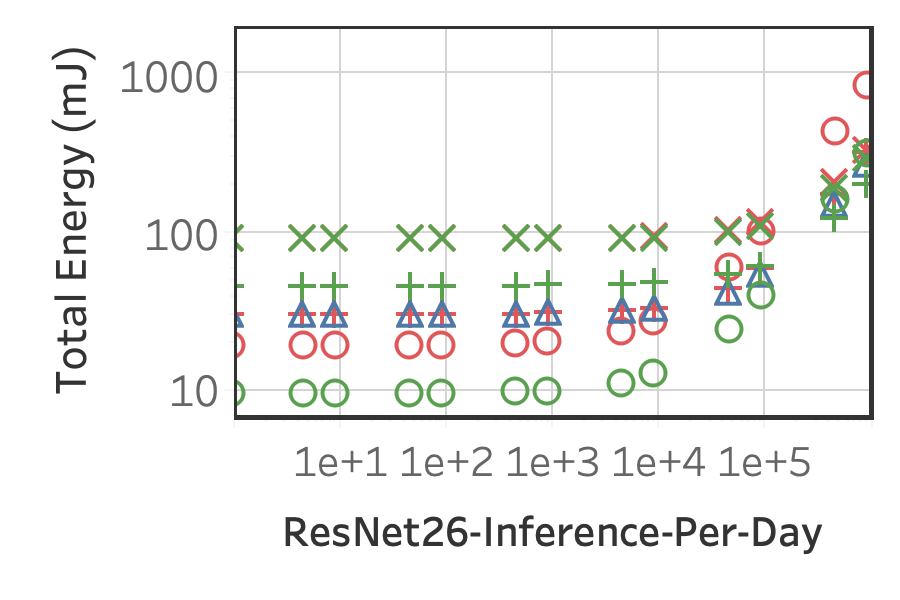}
    \includegraphics[width=0.47\textwidth]{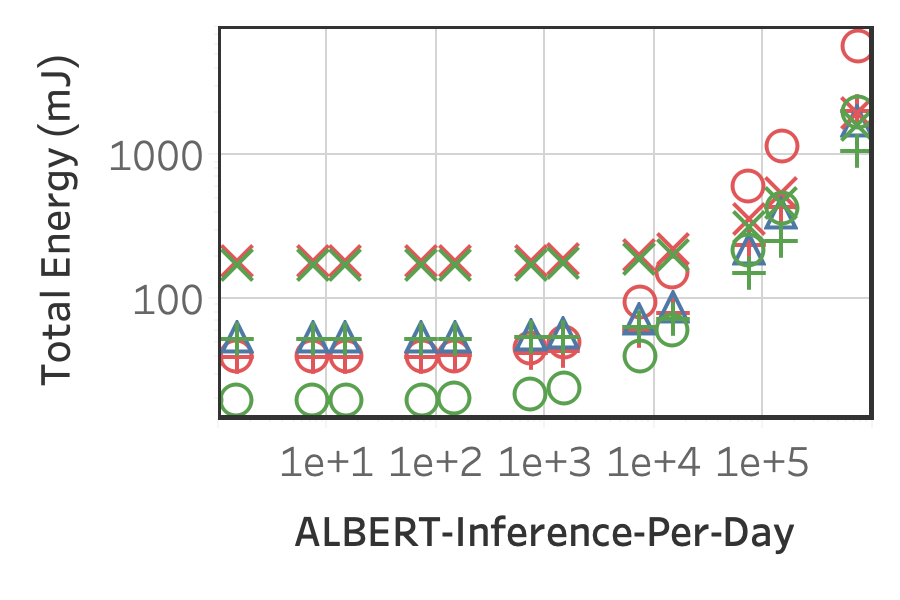}
    \caption{The eNVM storage solution (iso-capacity arrays provisioned per task, optimized for ReadEDP) that minimizes total memory energy consumption varies according to system wake-up frequency and DNN inference task; All solutions shown maintain application accuracy and a $<1s$ latency per inference.}
    \vspace{-15pt}
    \label{fig:intermittent}
\end{figure}

\begin{figure*}[t]
    \centering
    \includegraphics[width=0.98\textwidth]{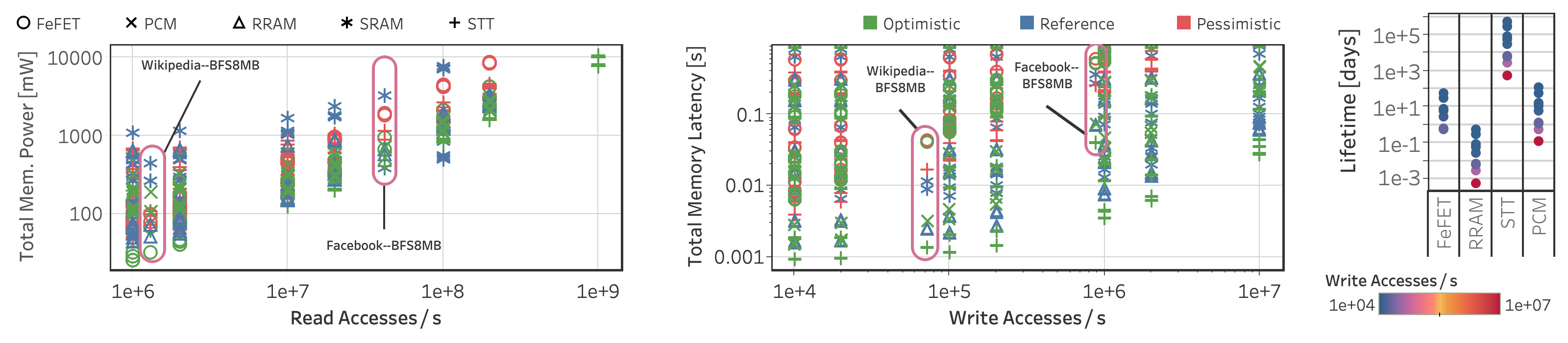}
    \vspace{-5pt}
    \caption{Memory power, latency, and projected lifetime for generic traffic patterns encompassing graph processing demands, including specific graph kernels as labeled. The lowest power solution depends on the expected read traffic. FeFET solutions fail to match SRAM performance. STT provides superior performance and memory lifetime.}
    \vspace{-10pt}
    \label{fig:graphs}
\end{figure*}

\subsubsection{Continuous Operation}

We consider the commonly-used and well-studied NVDLA \cite{nvdla_hotchips} as a base computing platform and compare its 2MB SRAM with iso-capacity eNVMs. 
We use the NVDLA performance model \cite{nvdla} to extract realistic memory access patterns and bandwidth requirements of the on-chip buffer.
More specifically, we evaluate the power and performance of accesses to on-chip memory storing ResNet26 weights for single-task image classification using the ImageNet dataset vs.\ multi-task image processing, comprising object detection, tracking, and classification, at a consistent frame rate of 60 frames-per-second, as is typical for HD video. 
We additionally consider the impact of storing activations in eNVM, but this ostensibly ignores endurance limitations.

%\subsubsection{Array Characteristics}
First, we observe the read and storage density characteristics for 2MB arrays using the cell-level tentpoles of several promising eNVM technology classes, as shown in Figure~\ref{fig:dnn_array} compared with SRAM.
Notice that read energy effectively divides arrays into two tiers. 
STT, PCM, and RRAM offer lower read energies and competitive read latencies vs,\ SRAM. 
In contrast, FeFET-based eNVMs suffer from higher read energies, but optimistic FeFET offers the highest storage density with low latency. 
At similar low latency, optimistic STT offers 6$\times$ higher density over SRAM. PCM and RRAM outperform SRAM in terms of both read latency and storage density. 
While such comparative insights can readily be extracted from this pair of plots, there are other important dimensions to also consider, and \NAME\ facilitates more comprehensive analyses that consider the impact of application priorities and system-level use cases on eNVM design decisions.

%Considering more dimensions of the design space,
Figure~\ref{fig:dnn_app} (left) summarizes total operating power (both dynamic access and leakage power) for the 2MB memory arrays characterized in Figure~\ref{fig:dnn_array} and accessed according to traffic patterns of different ResNet deployment scenarios, i.e., single- vs. multi-task and weights-only vs. storing both weights and activations. 
These results exclude eNVM candidates that cannot support 60 FPS operation nor maintain DNN accuracy targets. 
Recall \NAME\ includes fault injection wherein high eNVM fault rates can degrade model accuracy to unacceptable levels. 
While not explicitly shown here, \NAME\ exposes numerous additional interactions for users to probe and build intuition. 
For example, while total memory power increases as the number of accesses per frame increases to compute multiple tasks, the ratio of read-to-write traffic stays roughly the same. 
Hence, the relative power of eNVM arrays also remains similar.  
In particular, PCM, RRAM, and STT all offer over 4$\times$ reduction in total memory power over SRAM. 
One important reason for this is that SRAM leakage power will dominate compared to eNVM solutions, even under high traffic.
Of the energy-efficient solutions, STT offers best performance (lowest application latency per frame). 
In contrast, optimistic FeFET offers higher storage density while maintaining 60FPS and a 1.5-3$\times$ power advantage over SRAM.

\begin{table}[b!]
\caption{Summary of preferred eNVM under varying DNN use case, task, storage strategy, and optimization priority.}
\resizebox{0.99\textwidth}{!}{
\begin{tabular}{c|c|c|c|c|c}
\textbf{Use Case}                                          & \textbf{Inference Task}                           & \textbf{Data Storage} & \textbf{Priority} & \textbf{Opt. eNVM} & \textbf{Alt. eNVM} \\ \hline
\multirow{8}{*}{\begin{tabular}[c]{@{}c@{}}Continuous \\ (60IPS)\end{tabular}}              & \multirow{4}{*}{\begin{tabular}[c]{@{}c@{}}Single-Task \\ Image Classification\end{tabular}}             & \multirow{2}{*}{Weights Only}          & Low Power                              & PCM                                & PCM                                 \\ \cline{4-6} 
                                                           &                                                   &           & High Density                           & FeFET                              & CTT                                 \\ \cline{3-6} 
                                                           &                                                   & \multirow{2}{*}{Weights + Acts}        & Low Power                              & PCM                                & RRAM                                \\ \cline{4-6} 
                                                           &                                                   &        & High Density                           & STT                                & RRAM                                \\ \cline{2-6} 
                                                           & \multirow{4}{*}{\begin{tabular}[c]{@{}c@{}}Multi-Task \\ Image Processing\end{tabular}}      & \multirow{2}{*}{Weights Only}          & Low Power                              & PCM                                & RRAM                                \\ \cline{4-6} 
                                                           &                                                   &          & High Density                           & FeFET                              & CTT                                 \\ \cline{3-6} 
                                                           &                                                   & \multirow{2}{*}{Weights + Acts}        & Low Power                              & STT                                & RRAM                                \\ \cline{4-6} 
                                                           &                                                   &        & High Density                           & STT                                & RRAM                                \\ \hline
\multirow{10}{*}{\begin{tabular}[c]{@{}c@{}}Intermittent \\ (1IPS)\end{tabular}} & \multirow{2}{*}{\begin{tabular}[c]{@{}c@{}}Single-Task \\ Image Classification\end{tabular}}             & \multirow{2}{*}{Weights Only}          &           Low Energy/Inf                             &     RRAM                               &          RRAM                           \\ \cline{4-6} 
                                                           &                                                   &           &    High Density                                    &      FeFET                              &   CTT                                  \\ \cline{2-6} 
                                                           & \multirow{2}{*}{\begin{tabular}[c]{@{}c@{}}Multi-Task \\ Image Processing\end{tabular}}      & \multirow{2}{*}{Weights Only}          & Low Energy/Inf                                       &                            FeFET        &   FeFET                                  \\ \cline{4-6} 
                                                           &                                                   &          &    High Density                                    &     FeFET                               &    CTT                                 \\ \cline{2-6} 
                                                           & \multirow{4}{*}{\begin{tabular}[c]{@{}c@{}}Sentence Classification \\ (ALBERT)\end{tabular}} & \multirow{2}{*}{Embeddings Only}      &   Low Energy/Inf                            &   RRAM  &     RRAM      \\ \cline{4-6} 
                                                           &                                                   &        & High Density                  &   FeFET           &        CTT         \\ \cline{3-6} 
                                                           &                                                   & \multirow{2}{*}{All Weights}          &    Low Energy/Inf               &   STT          &     RRAM           \\ \cline{4-6} 
                                                           &                                                   &           & High Density                  &   FeFET           &       CTT         \\ \cline{2-6} 
                                                           & \multirow{2}{*}{\begin{tabular}[c]{@{}c@{}}Multi-Task NLP \\ (ALBERT)\end{tabular}}          & \multirow{2}{*}{All Weights}       & Low Energy/Inf                 &    STT           &       RRAM          \\ \cline{4-6} 
                                                           &                                                   &       &    High Density              &    FeFET          &      CTT          \\           
\end{tabular}
}
\label{tab:dnn}
\end{table}

\subsubsection{Intermittent Operation}

Let us now consider eNVM storage for two additional use cases that alter system-level optimization goals and corresponding eNVM selection, further highlighting the flexibility and ease of exploration the \NAME\ framework offers.
A major advantage of storing DNN weights in eNVMs is that non-volatility supports intermittent operation that powers off the accelerator between inferences. 
Using SRAMs would either consume leakage power to keep the weights memory powered on or consume power to restore the weights from off-chip memory, e.g., by incurring a latency and energy penalty by fetching from DRAM. 
In this use case, we provision monolithic eNVM storage to hold all DNN weights (e.g., up to 32MB for Natural Language Processing (NLP) tasks).
%, 4MB for NLP embeddings, and 16-32MB for multi-task NLP). 
For image processing, all weight memory accesses are to eNVM, eliminating the wake-up latency and power associated with loading parameters on-chip, in addition to reducing distance between compute system and higher-capacity memory.
%Previous work demonstrated that careful optimization between DNN properties and MLC eNVM storage can further increase storage density \cite{maxnvm}.

Figure \ref{fig:dnn_app} (right) compares the resulting memory-energy-per-inference across eNVMs for both single-task image classification and multi-task image processing, as determined by the total number of accesses to retrieve all DNN weights over the course of processing one input frame. 
The lowest-energy technology choice differs between the single vs. multi-task inference and, perhaps more interesting, both are eNVM candidates with \textit{lower} storage density (RRAM and pessimistic FeFET), as opposed to the highest density options (STT and optimistic FeFET), which hints at a cross-stack prioritization of read performance as opposed to cell size reduction, as probed further in Sec. \ref{subsec:area_eff}.
We repeat this study for single task vs. multi-task natural language processing using the ALBERT network, a relatively small-footprint, high-accuracy, transformer-based DNN \cite{albert}.

%\textcolor{red}{Add a note about CTT}
%For example, \textcolor{red}{full inference (weights + acts) leads to STT as a clear winner due both to energy efficiency and endurance}.
%Alternatively, \textcolor{red}{intermittent operation at a lower frame-rate in an embedded SoC (e.g., sensor node that wakes up and does batch size of 1 at a time) points towards RRAM and FeFET; can decide between those two proposals according to desired lifetime vs. technology maturity.}

%While Table \ref{tab:dnn} summarizes eNVM choices at a fixed rate of intermittent operation (1-inference-per-second), we find that frequency of wake-up and specific target application traffic patterns play a critical role in selecting preferred eNVM candidates, as described in Section \ref{subsec:intermittent}.
To further study this result, we dig into the implications of intermittent operation and compare the total energy versus the number of inferences per day, showing a continuum of wake-up frequency that may arise (e.g., deployed solar-powered agricultural sensors or satellites, or a voice-enabled assistant executing NLP tasks on wake-up). 
The left plot of Figure~\ref{fig:intermittent} shows total memory energy as a function of inferences per day for image classification. 
Here, total memory energy is presented as a proxy for device battery life. 
From the figure, we observe that when the number of inferences per day is sufficiently low (less than 1e5), optimistic FeFET yields the lowest energy. 
At higher wake-up frequency, optimistic STTs take over because of the relatively lower energy-per-access. 
Figure~\ref{fig:intermittent} (right) investigates the impact on an NLP workload. 
While results are similar, optimistic STT emerges as the best technology at lower inference rates (as compared to image classification), because ALBERT requires more computational power per inference than ResNet26. 

Table \ref{tab:dnn} summarizes the preferred eNVM technology across different use cases and tasks, with ``Opt. eNVM'' indicating the preferred choice under optimistic underlying cell characteristics and ``Alt. eNVM'' indicating the preferred technology assuming pessimistic assumptions and reference points, and table entries for intermittent operation are selected at a fixed wake-up rate. 
%As demonstrated in the domain-specific case of DNN inference under continuous vs. intermittent operation in Section \ref{subsec:dnn}, 
%As shown by these results, non-volatility of on-chip storage resources is particularly compelling for resource-constrained systems that experience intermittent power supply (e.g., deployed solar-powered agricultural sensors or satellites) or otherwise operate intermittently (e.g., a voice-enabled assistant executing NLP tasks on wake-up). 
Across a range of device wake-up frequencies and per-wake-up compute patterns, we observe that several eNVMs become compelling, and the preferred NVM choice for further investigation varies depending on both of these factors. 

%Figure~\ref{fig:intermittent} shows total energy as a function of inference per day for two workloads of interest: ResNet26 (image classification) and ALBERT (NLP). Here total energy is presented because it is the closest proxy we have for battery life. From the figure, we can observe that when the number of inferences per day is sufficiently low (less than 1e5 for ResNet and less than 1e4 for ALBERT), the opt. FeFET technology results in the lowest total energy. After those point, the opt. STT results in the lowest energy for both tasks. The difference in transition point between FeFET and STT for the two workloads is due to the fact that ALBERT has higher energy inferences since sentence classification is a particularly intensive task.

\begin{figure*}[t]
    \centering
    \includegraphics[width=0.98\textwidth]{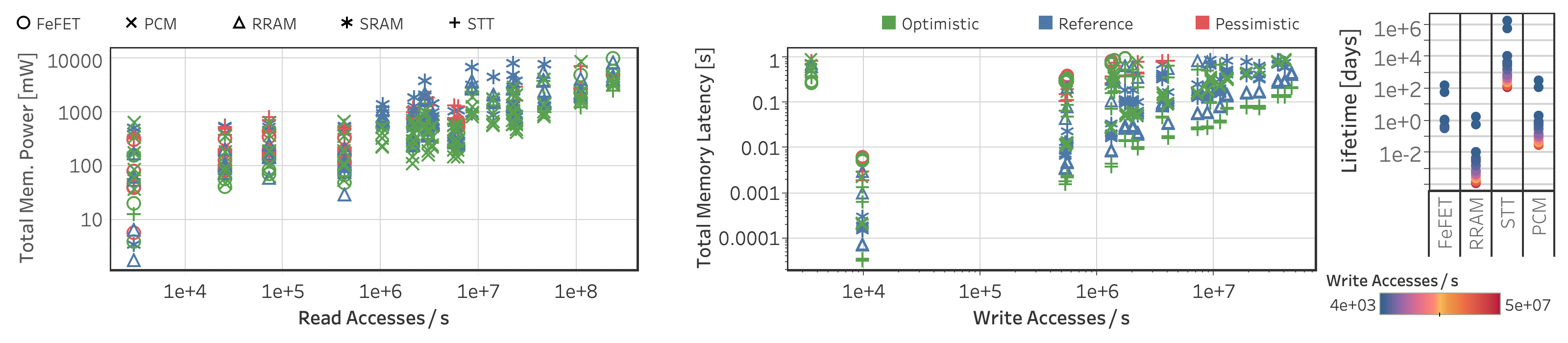}
    \caption{Memory operating power, latency, and projected lifetime under continuous operation across SPEC benchmark traffic to a 16MB LLC shows preferred eNVM depends on traffic demands and optimization goal. All solutions shown meet per-benchmark read/write demands. For high-traffic benchmarks, STT provides lowest power, lowest latency, and longest projected lifetime.}
    \vspace{-10pt}
    \label{fig:spec_power_16MB}
\end{figure*}

\vspace{-5pt}
\subsection{Enabling Efficient Graph Processing} \label{subsec:graphs}
Our second case study explores the potential benefits of using eNVMs for graph processing, which imposes an entirely different set of constraints in terms of memory read and write characteristics. 
Graph processing comprises many read-dominated tasks with less predictable data reuse than DNNs (e.g., search kernels), but still involves write traffic and, overall, is incredibly data-intensive in terms of memory bandwidth and capacity. 
As an initial exploration of compatibility and viability between graph processing workloads and eNVM storage solutions, we consider the total power and resulting memory lifetime per technology under generic traffic patterns covering the range of read and write bandwidths for critical graph tasks, as described in previous workload characterization efforts \cite{beamer_graphs}.
As a proof of concept in a specific system, we additionally evaluate eNVM storage solutions under access patterns for benchmarks executed on a domain-specific accelerator \cite{graph_accel}.

\subsubsection{Analysis for generic traffic patterns}
We consider different memories experiencing a range of generic traffic patterns representing graph processing kernels (i.e., read access rates from 1-10GB/s and write access rates from 1-100MB/s) \cite{beamer_graphs}.
\NAME\ provides a wide array of critical metrics to compare and user-configurable visualizations to extract important trends and limitations.
For example, in Figure~\ref{fig:graphs}, we choose to display total memory power against read traffic, as number of read accesses becomes a dominant factor in total power for read-dominated workloads, and  total memory latency against write traffic, as overall performance for several eNVMs is strongly determined by write traffic.

%\subsubsection{Power}
As shown by Figure~\ref{fig:graphs}, left, total memory power generally increases with read access rate and the lowest power solution depends on the application traffic load. 
For applications that exhibit fewer than 10$^7$ read accesses per second, optimistic FeFET is a clear winner, while pessimistic FeFET and RRAM are next best candidates. 
On the other hand, for higher rates of read traffic (e.g., $> 10^8$), optimistic STT is best. 
For mid-range read access rates, PCM and RRAM are also viable solutions sometimes offering the lowest power solution. 
However, this relationship alone does not dictate memory technology choice.
%\subsubsection{Latency}
A slightly different and more consistent story emerges when we analyze the impact of different eNVMs on overall memory latency (both read and write) versus write access rates, shown by the middle plot of Figure~\ref{fig:graphs}. 
While there is a clear preference for optimistic STT, RRAM and optimistic PCM are also worth considering. 
In contrast, most pessimistic eNVM technologies and all FeFET-based solutions are significantly inferior, even failing to match SRAM performance for many traffic patterns. 

When we additionally consider projected memory lifetime, STT emerges the clear winner overall. 
Note that the right chart of Figure~\ref{fig:graphs} plots the memory lifetime assuming continuous operation at a particular write access rate. 
Hence, the highest write traffic always yields the lowest lifetime. 
While RRAM seemed promising based on performance and power, it has the worst endurance and lowest lifetimes. 

\subsubsection{Analysis for domain-specific systems}
In addition to relying on generic traffic to represent the full range of expected load of graph processing, \NAME\ can also be leveraged to answer a more specific design question: For performance targets and traffic patterns to a specific storage resource in a graph processing accelerator system, which eNVMs offer compelling characteristics that warrant further investigation? 
To this end, Figure~\ref{fig:graphs} also includes points, identified in pink, corresponding to memory traffic to run breadth-first search on two different social network graphs \cite{snapnets}. 
Traffic patterns are extracted from throughput and accesses reported for the compute stream of a domain-specific graph processing accelerator utilizing an 8MB eDRAM scratchpad \cite{graph_accel}. 
In the baseline system, about $90\%$ of the energy is spent on the eDRAM scratchpad (not including DRAM controller energy), with an operating power of at least 3.1W at the 32nm process technology node as reported from Cacti \cite{graph_accel, cacti}. 
We analyze the benefits of replacing the 8MB eDRAM scratchpad with an iso-capacity eNVM array provisioned to meet the cited latency target (1.5ns).

If we exclude RRAM due to low lifetime projections, FeFET, PCM, and STT all offer significantly lower memory power (about 2-10$\times$ lower than SRAM) and even pessimistic STT offers consistent performance.
These observations, based on a realistic graph processing use case extracted from prior work, are consistent with the results generated using generic traffic patterns. 
Again, optimal technology choice depends on higher, system-level optimization goals, and \NAME\ provides critical insights in the presence or absence of a specific system solution and simulation results.

If the high-level goal is to maximize storage density, FeFET is highly attractive, but severely limited by poor write latency (unable to meet application latency expectations under the higher range of traffic patterns). 
Rather than prematurely eliminating FeFET, designers can leverage \NAME\ to study the impact of relaxing or adapting application targets or to explore co-design solutions that target improvements to the underlying technology (Sec.~\ref{subsec:bg_fefet}) or architecture (Sec.~\ref{subsec:write-buf}).

\vspace{-5pt}

\subsection{Non-Volatile LLC Solutions} \label{subsec:llc}

Improved density and energy efficiency could revolutionize general-purpose on-chip storage, and recent efforts have endeavored to replace high-performance memories, like caches, with eNVM-based alternatives~\cite{korgaonkar,hankin,deepnvm++}. 
However, caches must handle a large volume of writes depending on the application, so the achievable write latency and endurance per eNVM comes to the forefront of design considerations. 
%While the improved density and energy efficiency of eNVMs could revolutionize general-purpose on-chip storage, the open question of achievable endurance and write access characteristics per technology cannot be overlooked and needs to be a primary factor in determining a cache replacement. 
%This context has specific system constraints which on first glance are antithetical to the advantages of NVMs (superior read performance at cost of ). 

\begin{figure}[t]
    \centering
    \includegraphics[width=0.98\textwidth]{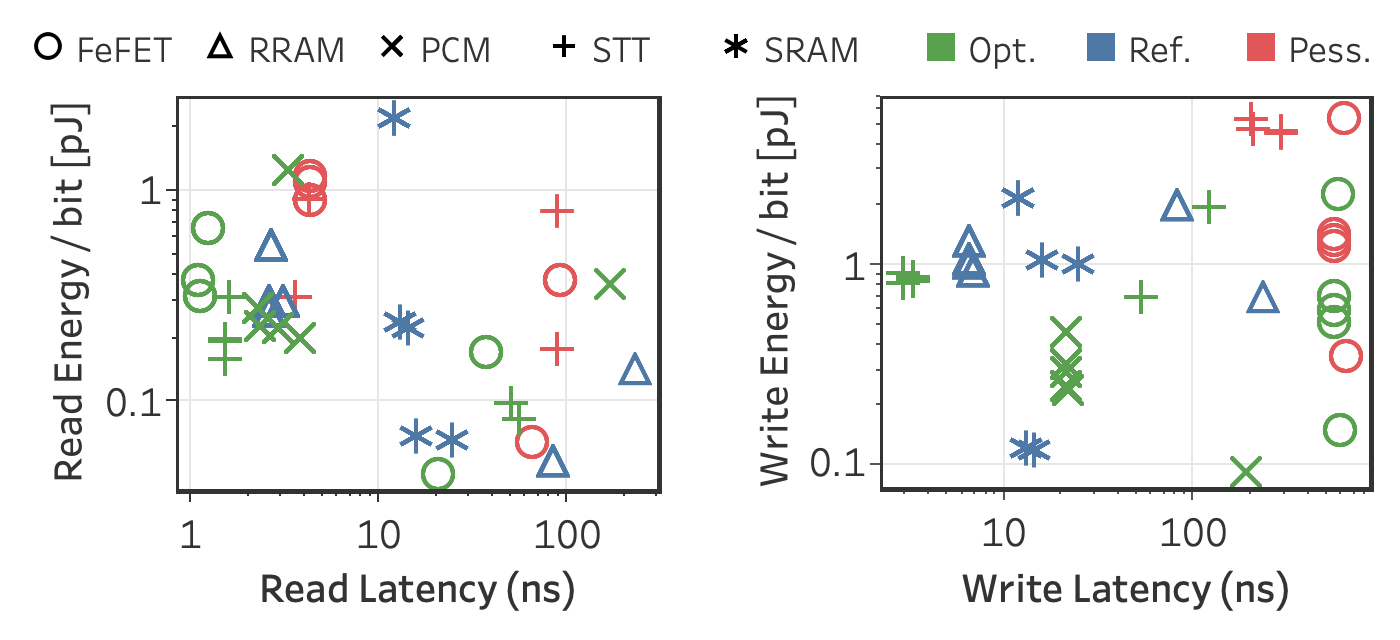}
    \caption{Array access characteristics in isolation for consideration of replacing (iso-capacity) a 16MB LLC.}
    \vspace{-10pt}
    \label{fig:array_comparison}
\end{figure}

In this study, we consider the last-level cache (LLC) of a high-performance desktop processor, similar to Intel's 14nm, 8-core Skylake. 
The memory hierarchy includes a private 32 KiB L1I\$; a private 32 KiB L1D\$; a private 512 KiB L2\$ (non-inclusive, write-back); and a shared ring 16MiB L3\$ with 64 B line, 16 ways (inclusive and write-back). 
The system includes DRAM with 2 channels, 8 B/cycle/channel, 42cycles + 51 ns latency.
%FIXME? I consolidated some system specs; either they need to be cut or put in a table?
%The memory hierarchy includes a private 32 KiB L1I\$ with 64 B line, 8 ways; a private 32 KiB L1D\$ with 64 B line and 8 ways with write-back policy; a private 512 KiB L2\$ with 1024 sets, 64 B line and 4-way (non-inclusive, write-back); and a shared ring 16MiB L3\$ with 64 B line, 16 ways (inclusive and write-back). 
%The system includes DRAM with 2 channels, 8 B/cycle/channel, 42cycles + 51 ns latency. 
Representative application behavior comes from SPECrate CPU2017 (integer and floating point), and we warm-up the cache for 500M instructions and simulate for 1-billion instructions in detail using the Sniper simulator \cite{SPEC_CPU2017, sniper}. 
This provides application modeling data for a 16MB LLC (e.g., reads, writes, execution time per benchmark) that are inputs to NVMExplorer (see Section~\ref{subsec:cross-stack}). 
%When replacing the last level cache with an NVM, we use an iso-capacity configuration. Another possibility is to fill up the same amount of physical area that an equivalent SRAM-based last level cache would consume (which would lead to significant capacity advantage for NVMs).

%\subsubsection{Array characteristics}
First we focus on the array characteristics of the different memory technologies in isolation, as shown in Figure~\ref{fig:array_comparison}. 
From the left plot, we note a competetive range of read energy and read latency does not reveal a clear winner. 
For example, if read energy per access is highest priority, FeFET, RRAM or even SRAM offer array configurations that trade access latency for energy efficient, while STT and optimistic FeFET offer pareto-optimal read characteristics. 
For writes (Figure~\ref{fig:array_comparison}, right), a PCM-based last level cache appears to minimize energy per access. 
On the other hand, only STT and RRAM are able to beat SRAM write latency. 
Again, we find array characteristics in isolation do not offer sufficient guidance to choose the best eNVM for LLC, and \NAME\ allows us to go further.

%In terms of storage density, STT shows over an order of magnitude reduction in area compared to SRAM (from about 10$mm^2$ to about 1$mm^2$ across optimization targets), though under pessimistic underlying cell characteristics, we note that PCM or RRAM offers higher density that STT. 

%\subsubsection{Power, Performance, and Lifetime}
%Now we start to focus on the effects of the actual application behavior on array characteristics in the context of the last level cache. 
%It is, therefore, important also consider system-level application behavior. 
Figure~\ref{fig:spec_power_16MB} shows the resulting power, performance, and lifetime when using different eNVMs as LLC and assuming memory traffic from SPEC2017 benchmarks. 
The leftmost figure shows total memory power versus read access rate, where each column of points corresponds to a particular benchmark traffic pattern. 
We again see that the lowest power eNVM solution depends on the traffic pattern. 
In broad terms, RRAM and FeFET fair better for lower read access rates while PCM is better for higher rates until STT emerges best for the highest rates. 
In terms of memory access latency with respect to write access rates, STT is usually the best choice, though arrays unable to meet application bandwidth are excluded. 
Lastly, the rightmost figure compares lifetimes across the eNVM technologies for a range of write access rates. 
Again, STT offers the best longevity on average. 
However, PCM and FeFET may warrant consideration for read-dominated workloads. 
RRAM, on the other hand, does not appear viable as an LLC. 

%Similar to Interestingly, the winning technology is dependent on the read accesses per second seen by the last level cache only when they are sufficiently low. Even more so is the fact that it is actually the pessimistic STT technology which outperforms all others for the case where the read accesses per second is 10$^{4}$. For higher values of read accesses per second, the optimistic version of STT is the clear winner. In terms of performance, total memory latency is consistently lowest for the optimistic STT regardless of number of write accesses per second.

%Now we get to one of the most important constraints for an architecture context with high, mixed access intensity: the last level cache. 
%In terms of lifetime, the rightmost plot compares resulting memory lifetime for the most promising last level cache replacement technologies given the observed application behavior. We note that while STT exhibits a more reasonable range of resulting memory lifetime across SPEC benchmarks, depending on the use case of your desktop machine, PCM may also be a compelling option (e.g., if read-dominated workloads such as xalancbmk dominate an end user's desktop machine, rather than more write-intensive scientific computing workloads.

%\textcolor{red}{TODO ``what if" points, iso area?}

%\subsection{Energy Harvesting, lightweight computation, intermittent, edge}

%\begin{figure}
%    \centering
%    \caption{Filtering previous case studies and adding application-level metrics, show same format and stats.}
%    \label{fig:ulp}
%\end{figure}

\vspace{-5pt}
\section{Co-Design Opportunities} \label{sec:codesign}
%\vspace{-10pt}

Exploration of the design space in Section~\ref{sec:case_studies} shows that no single eNVM technology is best. 
Rather, technology choice depends on the application and system-level targets. 
This also means there are ample co-design opportunities across the computing stack -- from devices to architecture. 
%\NAME\ encourages cross-stack co-design between architects and device designers. 
%By contextualizing and evaluating high-level implications of cell-level innovations as they emerge, one can identify what system-level opportunities are unlocked by that change. 

%%%%%%%%%%%%%%%%%%%
% CANDIDATE FOR CUT
%%%%%%%%%%%%%%%%%%%%
%As an example, Section~\ref{subsec:bg_fefet} motivates further development of back-gated FeFETs by revealing their promise for graph processing.
%Another opportunity is to relax certain cell-level parameters to better suit application or system priorities (Section~\ref{subsec:area_eff}). 
%\NAME\ can also guide designers to better understand which device-level characteristics most hinder adoption of a given eNVM in a given system context. 
%Consequently, Section \ref{subsec:mlc} motivates further investment in multi-level storage and probes their reliability under application accuracy bounds. 
%Section~\ref{subsec:write-buf} shows how \NAME\ can help designers study the impact of architectural techniques like write buffering to mask poor write characteristics that impede eNVM adoption. 
%%%%%%%%%%%%%%%%%%%

%\input{device_driven_opps}
%\input{arch_driven_opps}

\begin{figure}[t]
    \centering
    \includegraphics[width=0.98\textwidth]{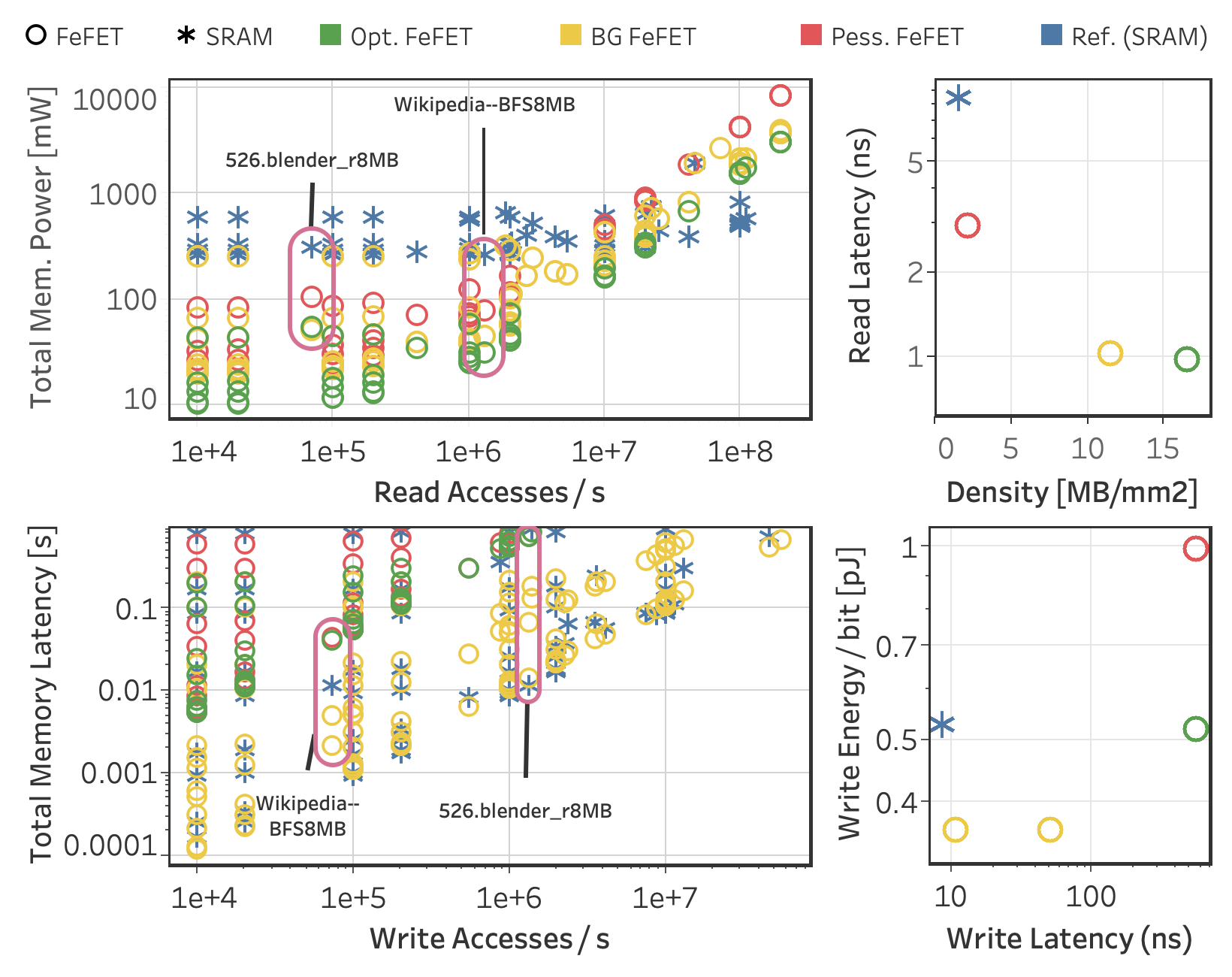}
    \caption{Back-gated (BG) FeFETs provide the high density and low operating power for example graph processing benchmarks with SRAM-comparable performance and begin to close the performance gap between non-BG FeFET and other memory technologies across SPEC2017 benchmarks.}
    \vspace{-10pt}
    \label{fig:bg_fefet}
\end{figure}

\subsection{Alternative FeFET fabrication choices unlock performant solutions for graph processing} \label{subsec:bg_fefet}

Previous FeFET-based device characterization and modeling efforts have exhibited write pulses on the order of $100ns$-$1\mu s$.
However, alternative FeFET fabrication strategies in early development stages, such as back-gated FeFETs \cite{intel_beol}, offer compelling potential advancements in write latency ($10ns$ programming pulse) and projected endurance ($10^{12}$). 
Section \ref{subsec:graphs} noted that the primary limition of FeFETs in the context of graph processing was an inability to meet the application latency targets under higher write traffic. 
Thus, using the underlying cell properties of back-gated FeFETs reported in \cite{intel_beol}, we can rapidly re-examine the viability of FeFET-based memory and probe whether this change could make a difference in the viability of FeFET-based memory for graph processing.

Figure \ref{fig:bg_fefet} shows the total memory power and total memory latency of an 8MB memory array of back-gated FeFETs (in yellow) compared to using previous FeFET standards (red, green) and SRAM (blue). 
We examine these metrics under a range of read and write traffic patterns which are inclusive of the graph benchmarks described in Section~\ref{subsec:graphs} and the SPEC benchmarks used in Section~\ref{subsec:llc}, but here showing access patterns for an 8MB capacity LLC. 
%Three different array organizations are tested which represent three different optimization targets: read latency, read energy, and read EDP. 
The underlying array-level characterization is shown in Figure \ref{fig:bg_fefet}, right. From the array characterization, we observe that the back-gated FeFETs show a slight increase in read energy per access and slight decrease in storage density compared to prior state-of-the-art cells. 
However, we observe that they enable comparable application latency to SRAM across a wide range of write traffic where previous FeFET versions fall short. 
Furthermore, back-gated FeFETs results in the lowest operating power over most of the range of read accesses per second, including for the example graph processing benchmark, Wikipedia--BFS8MB. 

Based on these observations, we posit that back-gated FeFET memory may close the performance gap between prior FeFETs and other memory technologies (including SRAM) and unlock additional application domains. 
\NAME's ability both to quickly and efficiently gauge the impact of cell-level innovations and to match emerging device designs to compelling use cases can enable productive future co-design collaborations. 
This feedback loop is mutually beneficial in providing direct motivation for further device development and encouraging system designers to integrate more energy-efficient, highly dense on-chip memory.

\begin{figure}[t]
    \centering
    \includegraphics[width=0.98\textwidth]{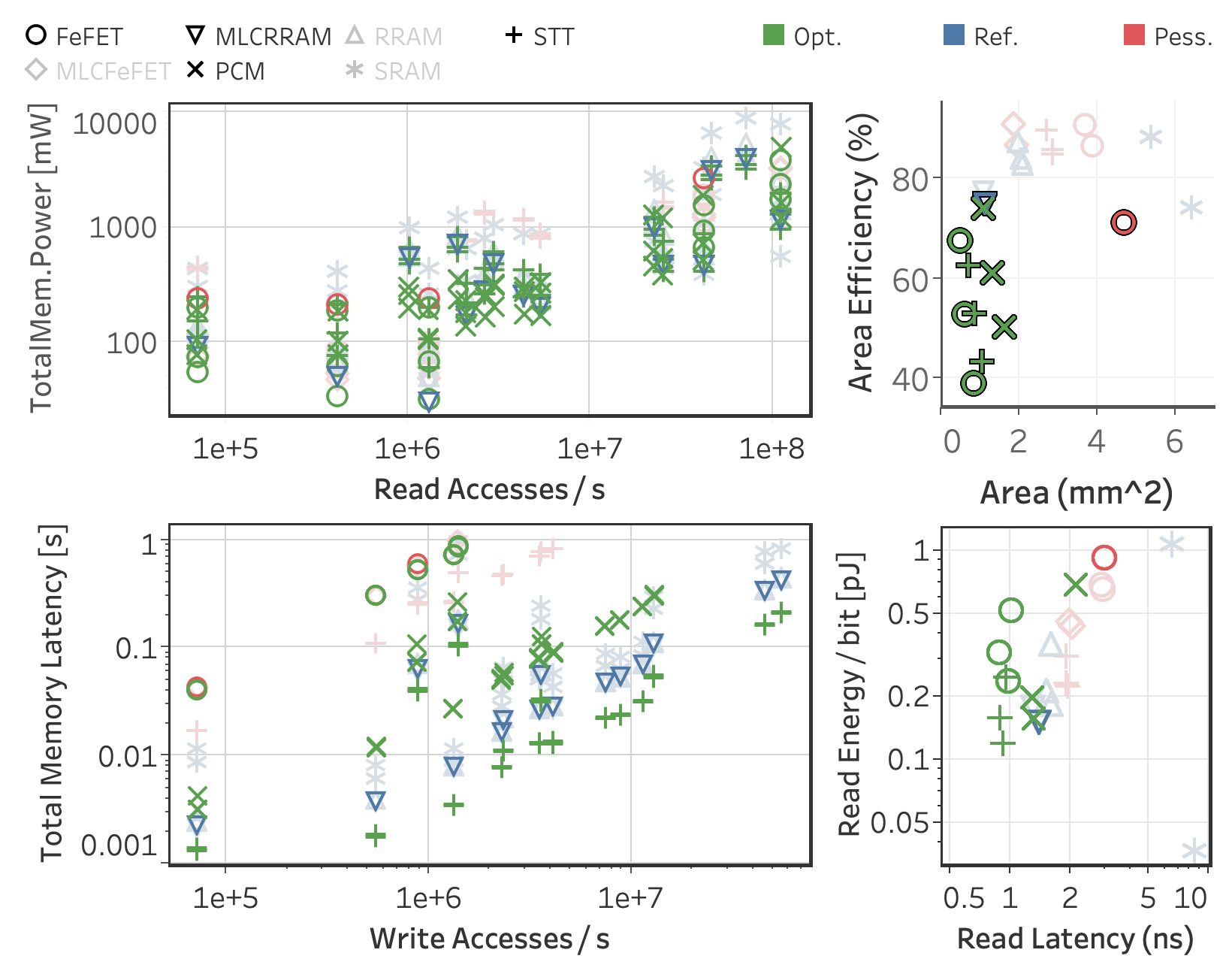}
    \caption{Results for 8MB arrays are filtered according to a maximum area efficiency (top right). Arrays with lower area efficiency are highlighted across all views and tend to result in low memory latency across many traffic scenarios.}
    \vspace{-10pt}
    \label{fig:area_eff}
\end{figure}

\vspace{-5pt}
\subsection{Trade area efficiency for performance} \label{subsec:area_eff}

One theme we can highlight across the architecture-driven case studies from Section~\ref{sec:case_studies} is that the subset of characterized results that exhibit lower area efficiency (i.e., internal array architectures that do less amortization of periphery and sensing overhead) also tend to result in lower total memory latency across many traffic scenarios. This is perhaps counter-intuitive given the effort spent in the devices community to manufacture very small cell sizes. We also note that in Figure \ref{fig:area_eff}, where such design points are highlighted across the plots, that slight advantages in terms of energy-per-access (e.g., Opt. STT and PCM compared to FeFET) tend to correlate to large total power advantages in high-traffic scenarios. As such, pointing out to device designers the greater relative impact of reduced energy per access rather than decreased cell size could usher in a more productive, product-ready set of eNVM technologies. 
Additionally, we observe that reducing energy per write access for STT and RRAM would drastically improve their relative power advantage for data-intensive applications, even at a cost of relatively lower area efficiency or storage density.

\subsection{Multi-Level Cell (MLC) advantages vary among eNVMs} \label{subsec:mlc}

\begin{figure}[t]
    \centering
    \includegraphics[width=0.98\textwidth]{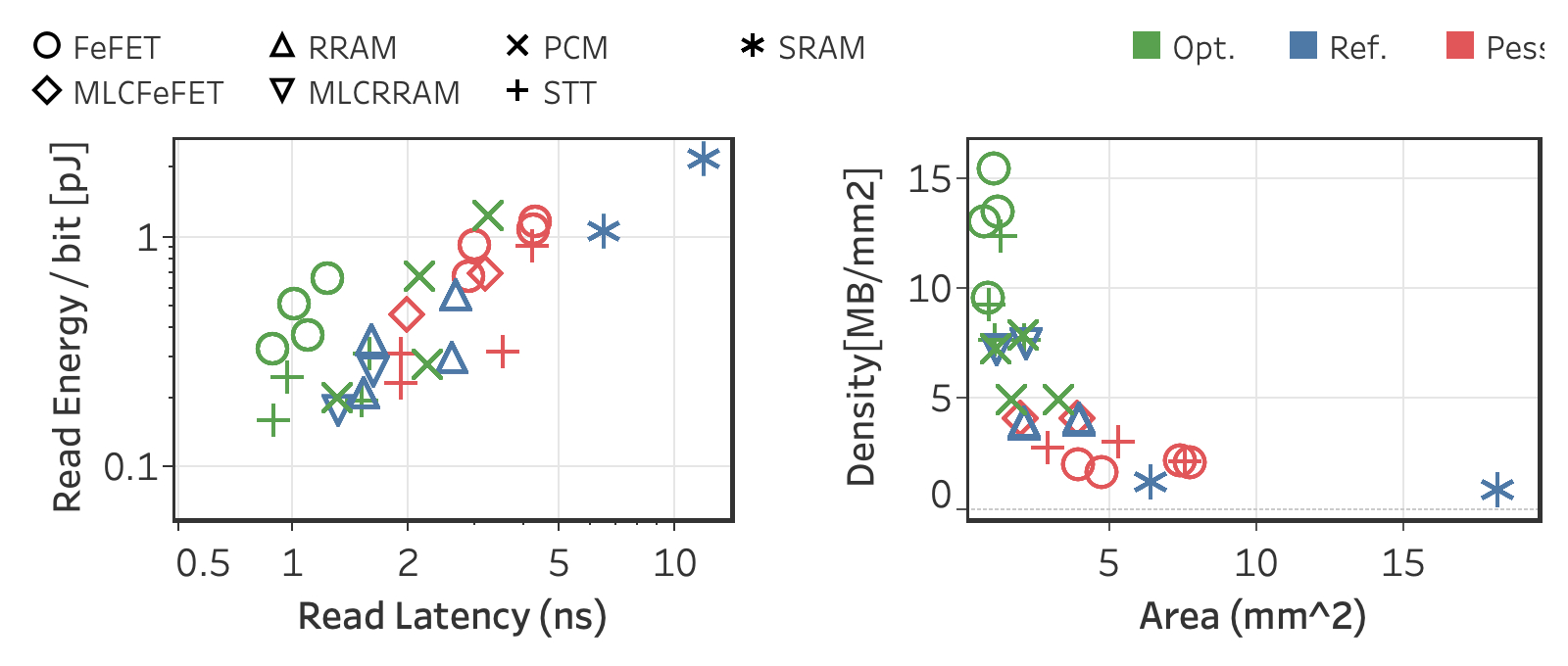}
    \caption{When we consider multi-level cells (MLC)s and filter out memory solutions that don't provide acceptable ResNet18 inference accuracy after fault injection, we note MLC RRAM is denser and more performant than SLC RRAM, while MLC FeFET is only sufficiently reliable for larger cell sizes (red).}
    \vspace{-12pt}
    \label{fig:mlc}
\end{figure}

While programming multiple bits per memory cell is an important strategy for increasing storage density across many eNVMs, previous work has revealed that MLC eNVMs may exhibit significantly higher fault rates that must be carefully considered in conjunction with application resilience \cite{maxnvm}. 
\NAME~enables efficient and broad probing of reliability vs. storage density by providing an application-agnostic fault injection tool and templates for technology-specific fault modes (Section~\ref{subsubsec:fault_modeling}).
To demonstrate, we quantify the application accuracy for ResNet18 image classification under weight storage in SLC vs. 2-bit MLC across multiple technologies for which there exists sufficient cell and circuit level data to produce detailed fault models. 
The density vs. reliability trade-off is distinct for each technology. 
For example, Figure \ref{fig:mlc} displays 8MB and 16MB characterized arrays, including 2-bit MLC RRAM and 2-bit MLC FeFET, filtered such that only those arrays meeting application latency requirements and maintaining image classification accuracy are included. 
Note that these results replicate previous efforts that indicate that image classification inference is robust to 2-bit MLC RRAM storage (we also verified this for CTT-based memories with fault modeling details provided in \cite{maxnvm, dac_ctt}), while we show that MLC FeFET devices only exhibit acceptable accuracy for larger cell sizes. 
This is because smaller FeFETs are more difficult to program reliably due to device-to-device variation \cite{fefet-arxiv}. 
%Portions of \NAME\ were leveraged to quantify cell- and circuit-level trade-offs specific to MLC FeFETs in greater depth to determine optimal cell provisioning and writing schemes for target applications \cite{fefet-arxiv}.

\subsection{Write buffering changes the performance landscape} 
\label{subsec:write-buf}

In conjunction with technology innovations to reduce write latency, adoption of a wider set of eNVMs in general-purpose computing contexts could be made possible by employing existing architectural techniques to mask poor write characteristics. 
For example, in an effort to extend memory lifetime and mask the performance impact of write access, a more performant technology (e.g., SRAM, or STT) could be employed as a write-buffer. 
Rather than employ a costly and engineering-intensive cycle-accurate simulator to gauge the impact of provisioning a write buffer, \NAME~enables an analytical study under user-specified traffic patterns to narrow the space of eNVMs worthy of further simulation and design effort. This approach answers high-level questions regarding whether write-buffering could make a difference in making additional eNVMs viable for applications with significant write traffic, and, if so, how much benefit would need to be extracted using the write buffer?

\begin{figure}[t]
    \centering
    \includegraphics[width=0.98\textwidth]{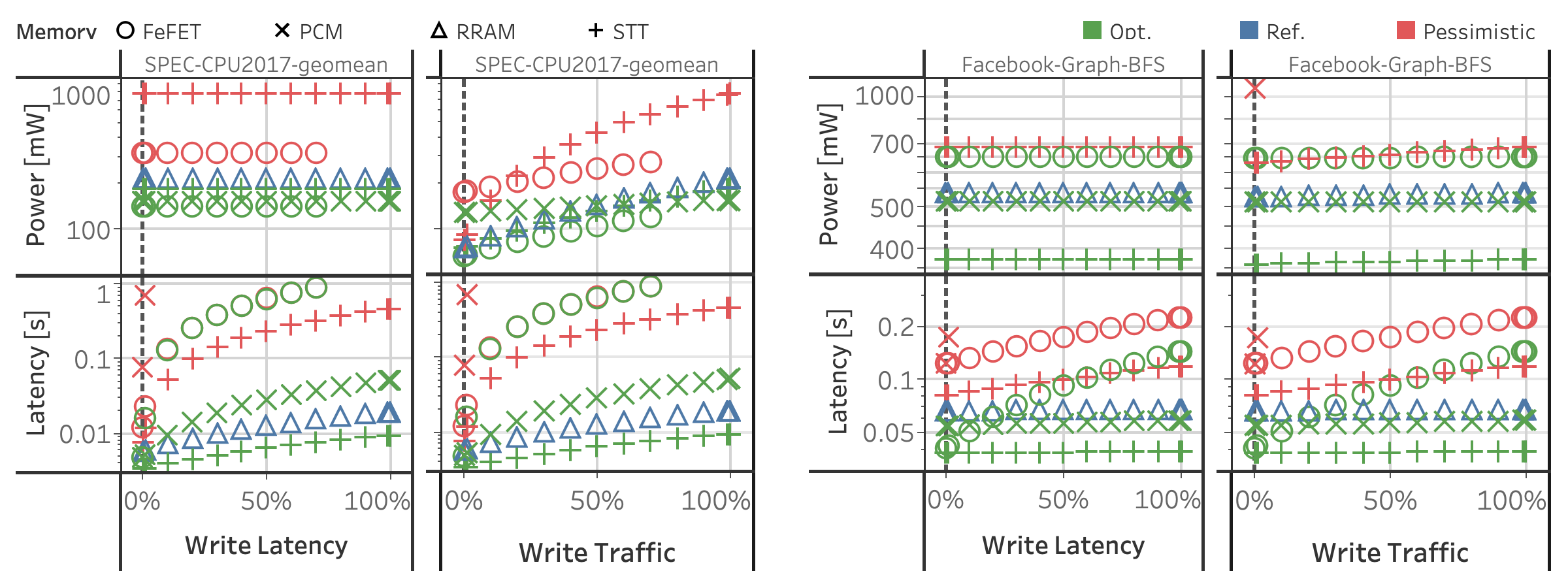}
    \caption{Masking write latency or reducing write traffic via introduction of a write caching scheme could enable a broader set of eNVM technologies.}
    %\vspace{-10pt}
    \label{fig:write_buffer}
\end{figure}

\begin{table}[b]
\caption{\NAME~leverages existing efforts by extending NVSim, while enabling cross-stack DSE across multiple use cases and domains, including more breadth than previous works, and providing a unified platform to explore and iterate design.} 
\begin{center} {
\resizebox{0.98\textwidth}{!}{

                                                  %     &                              & NVMain \cite{NVMain} & NVSim \cite{NVSim} & DESTINY \cite{DESTINY} & NeuroSim+ \cite{neurosim+} & IRDS Roadmap \cite{IRDS} & Memory Trends \cite{stanford_memory_trends} & \NAME~ \\ \hline 
\begin{tabular}{p{10mm}p{14mm}|p{8mm}|p{8mm}|p{10mm}|p{12mm}|p{12mm}|p{12mm}|p{12mm}||p{20mm}|}
\cline{3-10}
                                                                     &                         & \multicolumn{2}{c|}{Tech. Surveys}             & \multicolumn{2}{c|}{Array Simulators}         & \multicolumn{3}{c||}{Arch-Specific Frameworks} &  \multicolumn{1}{c|}{\textbf{This Work}  }                                           \\ \cline{3-10}
                                                                     &                         & IRDS \cite{IRDS}       & Mem. Trends \cite{stanford_memory_trends}     & NVSim\cite{NVSim}              & DESTINY\cite{DESTINY}           & Neuro- Sim+\cite{neurosim+}  & NVMain\cite{NVMain}  & Deep- NVM++\cite{deepnvm++}  & \textbf{NVMExplorer}                \\ \cline{3-9} \hline
\multicolumn{1}{|p{15mm}|}{}                                               & RRAM                    & \multicolumn{1}{c|}{\checkmark}                & \multicolumn{1}{c|}{\checkmark}                 & \multicolumn{1}{c|}{\checkmark}                 & \multicolumn{1}{c|}{\checkmark}                & \multicolumn{1}{c|}{\checkmark}         &               &            & \multicolumn{1}{c|}{\checkmark}                                        \\ \cline{2-10} 
\multicolumn{1}{|p{15mm}|}{}                                               & STT                &                          & \multicolumn{1}{c|}{\checkmark}                & \multicolumn{1}{c|}{\checkmark}                & \multicolumn{1}{c|}{\checkmark}                & \multicolumn{1}{c|}{\checkmark}        & \multicolumn{1}{c|}{\checkmark}     & \multicolumn{1}{c||}{\checkmark}  & \multicolumn{1}{c|}{\checkmark}                                       \\ \cline{2-10} 
\multicolumn{1}{|p{15mm}|}{}                                               & SOT                &                          & \multicolumn{1}{c|}{\checkmark}                &                          &                          &                  &               & \multicolumn{1}{c||}{\checkmark}  & \multicolumn{1}{c|}{\checkmark}                                        \\ \cline{2-10} 
\multicolumn{1}{|p{15mm}|}{}                                               & PCM                   &                          & \multicolumn{1}{c|}{\checkmark}                & \multicolumn{1}{c|}{\checkmark}                & \multicolumn{1}{c|}{\checkmark}                & \multicolumn{1}{c|}{\checkmark}        & \multicolumn{1}{c|}{\checkmark}    &            & \multicolumn{1}{c|}{\checkmark}                                       \\ \cline{2-10} 
\multicolumn{1}{|p{15mm}|}{}                                               & CTT                     &                          &                          &                          &                          &                  &               &            & \multicolumn{1}{c|}{\checkmark}                                      \\ \cline{2-10} 
\multicolumn{1}{|p{15mm}|}{}                                               & FeRAM                   & \multicolumn{1}{c|}{\checkmark}               & \multicolumn{1}{c|}{\checkmark}               &                          &                          &                  &               &            & \multicolumn{1}{c|}{\checkmark}                                       \\ \cline{2-10} 
\multicolumn{1}{|p{15mm}|}{\multirow{-7}{*}{NVM}}               & FeFET                   &                          & \multicolumn{1}{c|}{\checkmark}                &                          &                          & \multicolumn{1}{c|}{\checkmark}       &               &            & \multicolumn{1}{c|}{\checkmark}                                       \\ \hline \hline
\multicolumn{1}{|p{15mm}|}{}                                               & MLC & \cellcolor[HTML]{C0C0C0} & \cellcolor[HTML]{C0C0C0} &  &  & \multicolumn{1}{c|}{\checkmark}        &               &            & \multicolumn{1}{c|}{\checkmark}                                        \\ \cline{2-2} \cline{5-10}
\multicolumn{1}{|p{15mm}|}{\multirow{-2}{*}{Circuits}}                     & \begin{tabular}[l]{@{}l@{}}Fault \\ Modeling \end{tabular}         & \cellcolor[HTML]{C0C0C0} & \cellcolor[HTML]{C0C0C0} &  &  & \multicolumn{1}{c|}{\checkmark}        &               &            & \multicolumn{1}{c|}{\checkmark}                                       \\ \hline \hline
\multicolumn{2}{|p{30mm}|}{\multirow{4}{*}{\begin{tabular}[l]{@{}l@{}}Architectural \\ Simulator / \\ Use Case\end{tabular}}}                                        & \cellcolor[HTML]{C0C0C0} & \cellcolor[HTML]{C0C0C0} & \cellcolor[HTML]{C0C0C0} & \cellcolor[HTML]{C0C0C0} & Focus on PIM for DNNs     & gem5          & GPGPU- sim for DNNs  & Analytical; CPU, GPU, accelerator included \\ \hline \hline
\multicolumn{1}{|p{15mm}|}{\multirow{4}{*}{\begin{tabular}[p{15mm}]{@{}c@{}}App-Aware \\ Evaluation\end{tabular}}}                                               & Accuracy & \cellcolor[HTML]{C0C0C0} & \cellcolor[HTML]{C0C0C0} & \cellcolor[HTML]{C0C0C0} & \cellcolor[HTML]{C0C0C0} & \multicolumn{1}{c|}{\checkmark}        &               &            & \multicolumn{1}{c|}{\checkmark}                                        \\ \cline{2-2}  \cline{7-10}
\multicolumn{1}{|p{15mm}|}{}                                               & Memory Lifetime         & \cellcolor[HTML]{C0C0C0} & \cellcolor[HTML]{C0C0C0} & \cellcolor[HTML]{C0C0C0} & \cellcolor[HTML]{C0C0C0} &                  & \multicolumn{1}{c|}{\checkmark}     &            & \multicolumn{1}{c|}{\checkmark}                                        \\ \cline{2-2}  \cline{7-10}
\multicolumn{1}{|p{15mm}|}{}                                               & Operating Power         & \cellcolor[HTML]{C0C0C0} & \cellcolor[HTML]{C0C0C0} & \cellcolor[HTML]{C0C0C0} & \cellcolor[HTML]{C0C0C0} & \multicolumn{1}{c|}{\checkmark}        & \multicolumn{1}{c|}{\checkmark}     & \multicolumn{1}{c||}{\checkmark}  & \multicolumn{1}{c|}{\checkmark}                                        \\ \cline{2-2}  \cline{7-10}
\multicolumn{1}{|p{15mm}|}{} & Latency                 & \cellcolor[HTML]{C0C0C0} & \cellcolor[HTML]{C0C0C0} & \cellcolor[HTML]{C0C0C0} & \cellcolor[HTML]{C0C0C0} & \multicolumn{1}{c|}{\checkmark}        & \multicolumn{1}{c|}{\checkmark}     & \multicolumn{1}{c||}{\checkmark}  & \multicolumn{1}{c|}{\checkmark}                                       \\ \hline
\end{tabular}
}
} %\begin{tabular}[p{15mm}]{@{}c@{}}App-Aware \\ Evaluation\end{tabular}}

    \vspace{-10pt}
    \label{tab:scope} 
    \end{center}
\end{table}

For illustrative purposes, we consider a simple write cache that would hold write requests to the eNVM, write back to eNVM when the buffer is full, and allow in-place updates in the case of multiple writes to the same address before an update to eNVM.
Figure \ref{fig:write_buffer} shows the results for this study for SPEC2017 and Facebook-Graph-BFS. 
Just buffering the writes will mask the effective write latency experienced by the system, while a write cache that allows updates could additionally reduce traffic and extend lifetime.
In particular, we look at the effects of masking write latency and reducing write traffic on total memory latency and power. 
We observe that for Facebook-Graph-BFS, if the write traffic load is reduced by at least half, FeFET emerges as a performant option, while STT remains the lowest power solution for this particularly high-traffic workload. 
STT and RRAM are still the optimal technology choices for SPEC2017 in terms of performance, but write-buffering could empower FeFETs as a lower-power alternative if latency could be masked or write traffic to the eNVM could be reduced by at least 25\%.

\vspace{-5pt}
\section{Related Work} \label{sec:relatedwork}

%distinguish from prior work using Table;
%describe surveys, then array-level, then arch-specific, then repeat contributions
Previous work in evaluating eNVMs can be characterized as either focusing on device- and array-level evaluations, or providing in-depth cross-stack analysis for a particular combination of eNVM and application target. 
In Table~\ref{tab:scope}, we codify the key differences between \NAME\ and related works. Survey works such as the Stanford Memory Trends \cite{stanford_memory_trends} maintain a list of key parameters, like storage capacity and write energy, while previously validated array-level characterization tools, such as NVSim  \cite{NVSim}, characterize timing, energy, and area of eNVM-based memory structures. 
DESTINY \cite{DESTINY} modifies NVSim to evaluate 3D integration and could be similarly extended and used as a back-end characterization tool for \NAME.

To evaluate eNVMs in a system setting, prior work typically integrates NVSim with a system simulator. 
DeepNVM/DeepNVM++ \cite{DeepNVM,deepnvm++} enables cross-layer modeling, optimization, and design space exploration of MRAM-based technologies in the context of GPU cache for DNNs using GPGPUSim.
NVMain \cite{NVMain} enables evaluation of eNVM-based main memory using gem5. 
NeuroSim+ \cite{neurosim+} focuses on evaluation of processing-in-memory for DNN inference and training. 
While these frameworks are great examples of domain-specific explorations and evaluations, NVMExplorer can evaluate a variety of system and application domains, in addition to offering reliability analysis, additional metrics such as memory lifetime, and a database of technology cell characteristics and configurable device parameters.

%\vspace{-10pt}

%Existing works such as these provide limited or otherwise domain-specific design space exploration frameworks. 
In contrast, \NAME\ offers more breadth by including application-, system-, and device-level considerations, and accommodating a wider range of devices without requiring a separate system simulator. Additionally, \NAME\ offers a broad range of evaluations, including fault modeling and reliability studies. It is built for ease of navigation and fluidity, and it exposes the unique cross-stack trade-offs among application characteristics, system constraints, and circuit and device level innovations in a user-friendly configuration interface and companion data visualization interface. 
%Presented case studies with \NAME\ have additionally integrated traffic patterns extracted from system simulators and previous publications to gauge the viability of eNVMs from a variety of system and application perspectives.
By integrating these components, \NAME\ additionally provides a platform for architects and device designers to perform co-design evaluations required for the advancement of technologically-heterogeneous memory systems. 

%The time required to sift through non-volatile memory roadmaps, learn the growing number of NVM simulators, and piece together toolflows to answer questions about specific architectures and applications is throttling the pace of co-design between architects and device designers and inhibiting the advancement of technologically-heterogeneous memory systems.

%In contrast to these works, \NAME's cross-stack and analytical approach provides efficient evaluation of critical application-aware metrics such as application accuracy and resulting memory lifetime with sufficient fidelity to guide future design studies.

%Furthermore, data from NVSim will then need to be plugged into an architecture simulator to actually see the effects of a new memory technology on a system. Some work has attempted to do this. 

%\begin{figure*}
%    \centering
%    \includegraphics[width=1.0\hsize]{figures/Limit %Studies Comparisons.png}
%    \caption{Alternative overview of \NAME.}
%    \label{fig:cycle_framework}
%\end{figure*}
\vspace{-5pt}
\section{Conclusion} \label{sec:conclusion}
 
%The energy-efficiency and scalability of traditional on-chip memory are critical limitations in a wide set of important computing systems today. 
Next-generation on-chip memory will need to push the boundaries of efficiency and density, and a diverse set of embedded non-volatile memory (eNVM) technologies have compelling characteristics to address these limitations.  
\NAME\ provides architects the flexibility to explore and compare these storage solutions under realistic constraints. 
%As a demonstration of \NAME's capabilities, we evaluate and compare eNVM solutions for DNN inference tasks, graph processing, and general-purpose computing scenarios. 
%We find that depending on system optimization goals and application properties, each eNVM emerges as a compelling candidate in at least one critical computing context, and there are key limiting characteristics both at the application level and the cell level. 
%This finding suggests the existence of cross-stack optimization opportunities, and \NAME\ empowers efficient and informative co-design studies such as alternative FeFET fabrication strategies to improve write access and enable support for graph processing or incorporating write caching to change the relative performance and power benefits of various eNVM solutions. 
\NAME~is open source, with interactive data visualizations freely available online, which we hope will unlock the potential of eNVMs in a broad range of systems.

\newpage

%%%%%%% AE APPENDIX %%%%%%%
\appendix
\section{Artifact Appendix}

%%%%%%%%%%%%%%%%%%%%%%%%%%%%%%%%%%%%%%%%%%%%%%%%%%%%%%%%%%%%%%%%%%%%%
\vspace{-5pt}

\subsection{Abstract}

NVMExplorer is an open-source framework for modeling, evaluating, and comparing embedded non-volatile memory solutions under different application and system-level properties and constraints.
NVMExplorer's code base includes (1) a python-based user interface for configuring and running design sweeps, (2) a modified and extended version of NVSim for memory array characterization, (3) a python-based application-level fault injection tool with a stand-alone interface, (4) scripts to both generate and parse associated configuration and output files from memory characterization, and (5) an analytical model extrapolates array-level data according to user-input application and system properties and constraints.
This release also includes (1) a per-technology database of properties extracted from paper survey of IEDM, VLSI, and ISSCC 2016-2020, (2) application characteristics for the workloads in our submission, including graph search, neural networks, and SPEC CPU2017, (3) fault model characteristics and data format transformations for fault injection studies, and (4) sample configuration files and customized cell-level characteristics.

NVMExplorer was developed and validated on both Mac and Ubuntu Linux systems, with successful configuration and some tests also on Windows.
Support for more advanced technology nodes and alternative memory characterization backends is under development.

\vspace{-5pt}

\subsection{Artifact check-list (meta-information)}

%{\em Obligatory. Use just a few informal keywords in all fields applicable to your artifacts
%and remove the rest. This information is needed to find appropriate reviewers and gradually 
%unify artifact meta information in Digital Libraries.}

{\small
\begin{itemize}
%  \item {\bf Algorithm: }
%  \item {\bf Program: }
  \item {\bf Compilation:
  \begin{verbatim}$ cd nvmexplorer_src/nvsim_src
$ make\end{verbatim}}
%  \item {\bf Transformations: }
%  \item {\bf Binary: }
%  \item {\bf Model: }
%  \item {\bf Data set: }
%  \item {\bf Run-time environment: }
%  \item {\bf Hardware: }
%  \item {\bf Run-time state: }
  \item {\bf Execution: \begin{verbatim}$ python run.py config/main_dnn_study.json\end{verbatim}}
%  \item {\bf Metrics: }
  \item {\bf Output: \begin{verbatim}output/results/[eNVM]_1BPC-combined.csv\end{verbatim}}
  \item {\bf How much disk space required (approximately)?: \normalfont{37MB}}
  \item {\bf Preparation Time?: \normalfont{Less than 1 hour.}}
  \item {\bf Execution Time (for this artifact)?: \normalfont{about 1 hour (desktop CPU).}}
  \item {\bf Publicly available?: \normalfont{Yes}}
  \item {\bf Code licenses (if publicly available)?: \normalfont{MIT}}
  \item {\bf Archived (provide DOI)?: \url{https://zenodo.org/badge/latestdoi/375786583} }
  
  %\item{
  %\includesvg{zenodo_v0_badge.svg} }
  
  %\externalfigure[https://zenodo.org/badge/doi/10.5281/zenodo.5684832.svg]
\end{itemize}
}

%%%%%%%%%%%%%%%%%%%%%%%%%%%%%%%%%%%%%%%%%%%%%%%%%%%%%%%%%%%%%%%%%%%%%

\vspace{-5pt}

\subsection{Description}

\subsubsection{How to access}

[\href{https://zenodo.org/badge/latestdoi/375786583}{Zenodo}] for current release, [\href{https://github.com/lpentecost/NVMExplorer}{Github}] for up-to-date and development versions.

\subsubsection{Hardware dependencies}

Tested on a selection of laptop and desktop setups; no specific HW dependencies required.

\subsubsection{Software dependencies}

Python 3.8; pandas, numpy, (optional) pytorch for fault injection experiments; gcc

\subsubsection{Data sets} 

Please see provided workload characterization results and default configuration settings in the following paths from the main NVMExplorer repository: \begin{verbatim} config/README.md\end{verbatim} \begin{verbatim} data/workload_data\end{verbatim} \begin{verbatim} output/NVM_data\end{verbatim}

%\subsubsection{Models}

%%%%%%%%%%%%%%%%%%%%%%%%%%%%%%%%%%%%%%%%%%%%%%%%%%%%%%%%%%%%%%%%%%%%%

\vspace{-5pt}

\subsection{Installation}

\begin{verbatim}$ git clone --recurse-submodules
https://github.com/lpentecost/NVMExplorer
$ cd nvmexplorer_src/nvsim_src
$ make\end{verbatim}

Prior to running NVMExplorer, please verify you are using Python 3.8 and have the pandas and numpy packages available.

%%%%%%%%%%%%%%%%%%%%%%%%%%%%%%%%%%%%%%%%%%%%%%%%%%%%%%%%%%%%%%%%%%%%%

\vspace{-5pt}

\subsection{Experiment workflow}

The general usage of NVMExplorer is via passing a JSON config file that specifies your desired design sweep to run.py:

\begin{verbatim}python run.py config/[config name].json\end{verbatim}

For example, to generate the DNN-focused case study in Section 4.1, you can run:

\begin{verbatim}python run.py config/main_dnn_study.json\end{verbatim}

and verify that the per-eNVM-technology CSV outputs generated are consistent with those provided in 
\begin{verbatim}AE_dnn_output/[eNVM]_1BPC-combined.csv\end{verbatim}

%%%%%%%%%%%%%%%%%%%%%%%%%%%%%%%%%%%%%%%%%%%%%%%%%%%%%%%%%%%%%%%%%%%%%
%\subsection{Evaluation and expected results}
%
%{\em Obligatory}

%%%%%%%%%%%%%%%%%%%%%%%%%%%%%%%%%%%%%%%%%%%%%%%%%%%%%%%%%%%%%%%%%%%%%
%\subsection{Experiment customization}
%\begin{verbatim}config/README.md
%https://nvmexplorer.seas.harvard.edu/docs/
%  _build/html/index.html\end{verbatim}
%%%%%%%%%%%%%%%%%%%%%%%%%%%%%%%%%%%%%%%%%%%%%%%%%%%%%%%%%%%%%%%%%%%%%
%\subsection{Notes}

%%%%%%%%%%%%%%%%%%%%%%%%%%%%%%%%%%%%%%%%%%%%%%%%%%%%%%%%%%%%%%%%%%%%%
%\subsection{Methodology}
%
%Submission, reviewing and badging methodology:
%
%\begin{itemize}
%  \item \url{https://www.acm.org/publications/policies/artifact-review-badging}
%  \item \url{http://cTuning.org/ae/submission-20201122.html}
%  \item \url{http://cTuning.org/ae/reviewing-20201122.html}
%\end{itemize}

\newpage

%%%%%%%%% -- BIB STYLE AND FILE -- %%%%%%%%
\bibliographystyle{IEEEtranS}
\bibliography{refs}
%%%%%%%%%%%%%%%%%%%%%%%%%%%%%%%%%%%%

\end{document}